\documentstyle[11pt,psfig]{article}
\def\be{\begin{equation}}
\def\ee{\end{equation}}
\begin{document}
\title{Study of Charmonia near the deconfining transition on an anisotropic
lattice with O(a) improved quark action}

\author{T. Umeda, R. Katayama, O. Miyamura\\
\\
Department of Physics, Faculty of Science, Hiroshima University,\\
Kagamiyama 1-3-1, Higashi-hiroshima, 739-8526, JAPAN,\\
\\
H. Matsufuru\\
\\
Research Center for Nuclear Physics, Osaka University,\\
Mihogaoka 10-1, Ibaraki  567-0047, JAPAN}

\date{\today}
\maketitle

\begin{abstract}
We study hadron properties near the deconfining transition
in the quenched lattice QCD simulation.
This paper focuses on the heavy quarkonium states, such as
$J/\psi$ meson.
In order to treat heavy quarks at $T>0$, we adopt the $O(a)$ improved
Wilson action on anisotropic lattice.
We discuss $c\bar{c}$ bound state observing the wave function
and compare the meson correlators at above and below $T_c$.
Although we find a large change of correlator near the $T_c$, the strong 
spatial correlation which is almost the same as confinement phase
survives even $T\sim 1.5T_c$.
\end{abstract}

\section{Introduction}
\label{sec:intro}
It is generally believed that the Quantum Choromodynamics (QCD)
exhibits a phase transition at some temperature $T_c$,
and quarks and gluons confined in the low temperature phase
are liberated to form the ``quark gluon plasma''.
At the beginning of 2000, CERN reported that the QGP state
had been created in the heavy ion collision experiment\cite{c1}.
In these experiments, the $J/\psi$ suppression\cite{c2} is regarded
as the key signal of QGP formation.
Since c-quark is heavy ( $m_c=1.15\sim 1.35$ GeV\cite{c3}) $c\bar{c}$
pair are hard to be generated except by the primary collisions of
nucleons in the high energy heavy ion, i.e. the $J/\psi$ is not created with
the thermal effects after the QGP formation.
Therefor it is expected that the effects of deconfining clearly
appear in this signal.
Further investigation will be performed in RHIC project at BNL.

On the theoretical side, in spite of the various 
approaches\cite{c4,c5,c6}, we are still far from the definite
understanding of hadron properties near the transition and
the fate of the hadronic states in the plasma phase.
Since the phase transition changes the relevant degree of freedom of the
system, the model approach which a priori assumes dynamical degrees of the
system is difficult to treat physics near the phase transition.
We investigate these problems using lattice QCD, which enables
us to incorporate the nonperturbative effect of QCD from the first
principle.

For a long time, in the lattice QCD simulations, hadron masses at finite
temperature have been argued with the spatial correlation
(screening mass)\cite{c7}.
On the other hand, the study of the temporal correlation, which is
related to the pole mass, has been started rather recently\cite{c8}.

The previous work\cite{c8} studied hadron properties near the $T_c$
based on light quarks.
They caught a sign of chiral symmetry restoration above $T_c$, 
and a change of correlators in temporal and spatial direction near the
$T_c$.
From the discussion of wave function, however, they find the same strong
spatial correlation at $T>T_c$ as that of below $T_c$, in other words
hadronic mode at a long range survives in the deconfinement phase. 
At present, there is no well-established way of lattice simulations to
attack the spectroscopy at $T>0$. However, our recent work seems one of
the best approaches to the problem.
It is interesting to apply this analysis to the heavy quarks.
In this paper we focus on the heavy quarkonium state,
which plays an important role as a signal of the quark gluon
plasma formation\cite{c2,c9}.

Our goal of the investigation is the prediction about $J/\psi$ suppression
or mass shift of charmonium\cite{c9} as signals of QGP formation
and the understanding of hadron properties and nature of QGP phase near
the transition.

\bigskip

For the study of charmonium physics at $T>0$ on a lattice, there are
several problems. Then we classify these problems into two
category, and discuss them individually.
\begin{description}
 \item[(i)] Precise calculation of temporal correlator of charmonium at
	    $T>0$ 
 \item[(ii)] Extraction of physical properties of charmonium from the 
	    correlators 
\end{description}
Firstly we consider the former one.
In the lattice QCD simulation at $T>0$,
we set a temporal lattice extent to $1/T$.
At high temperature, one needs the large lattice cutoff
to work with the sufficient degrees of freedom in the 
temporal direction.
In order to obtain the detailed information of temporal 
meson correlators at $T>0$, a high resolution in temporal
direction is needed.
The large lattice cutoff is also necessary to study a correlator
of meson with the heavy quarks because of its rapid decreasing behavior.
If one tries to overcome these difficulties with straightforward way,
the tremendous large computational power is necessary.
In order to get the sufficiently fine resolution
with limited computer resources, we use the anisotropic lattice,
which has a finer temporal lattice spacing $a_{\tau}$ than
the spatial one $a_{\sigma}$.

\bigskip

In this work we adopt the same strategy as the previous work\cite{c8} 
which was tractable to analyze light hadrons at $T>0$.
This strategy is as follows.
Firstly the mesonic operator are defined, then we observe its
correlator and wave function (in the Coulomb gauge) at $T=0$. 
Next we investigate how they are affected by the temperature.
In order to investigate the temperature effects for the state of
interest, for example a ground and excited state of charmonium,
we have to make the good mesonic operator which has the large overlap
with the state. 
Because we should compare correlators at $T \geq 0$ within shorter
temporal lattice extent.
The wave function gives hints for existence of mesonic state at
$T>0$. Especially we are interested in that of deconfinement phase. 
In the case of charmonium this wave function is an important quantity
concerning the $J/\psi$ suppression.

We prepare two sets of gauge configuration whose lattice spacing are
different.
We control the temperature by changing temporal lattice extent $N_t$.
Then above investigation is performed on these configurations.

This paper is organized as follows.
In the next section we define the quark action on anisotropic
lattice and discuss the dispersion relation of free quark and
calibration of quark field. Sect.~\ref{sec:setup}
is the preparation for
study of charmonium correlator. Here various parameters of gauge
configurations are determined and the calibration are performed. 
In Sect.~\ref{sec:zero} 
we report the charmonium spectroscopy and construction
of optimized operator using variational analysis at $T=0$.
Sect.~\ref{sec:ft} 
describes correlators at $T>0$ and the measurement of the wave function  
and compare these with results at $T=0$.
The last section is the conclusion and discussion. 

\section{Quark action on the anisotropic lattice}
\subsection{Quark action}
To treat the quark field on the lattice, we adopt the $O(a)$ improved
Wilson quark formulation.
To construct the quark action on the anisotropic lattice,
we follow El-Khadra, Kronfeld and Mackenzie\cite{c10},
for the following advantages.
They expand the lattice Hamiltonian in the power of $a$, and
determine the coefficient of each operator by matching the lattice 
Hamiltonian with the the continuum one except the redundant operators.
The resultant action takes the same form as the clover quark
action\cite{c11} in the limit of $m\rightarrow 0$.
On the other hand, in the heavy quark mass region
($m_q \gg \Lambda_{QCD}$), the effective-theoretical treatment of
the quark action enables us to use it for such a quark on a lattice
of moderate cutoff.
Since the full quark mass dependence is incorporated, the same form
also covers the intermediate quark mass region, and then
the small and the large mass regions are smoothly connected.
Although our main target in this paper is the charm quark, it would be
useful to take the other mass region into account for the future
applications.
In addition to these advantages, 
their argument is naturally in accord with the anisotropic lattice.
They introduced the different hopping parameters for the spatial and
the temporal directions, and proposed to tune them so that the rest and
the kinetic mass take the same value.
Such a treatment is inevitably required on the anisotropic lattice
to assure that the anisotropy of the quark and the gauge fields
coincide, especially if one employ the dispersion relation for
the definition of the quark field anisotropy.

On the anisotropic lattice, the quark action takes almost the same form
as in the Ref.~\cite{c10} 
 \footnote{
 The notation in this paper is slightly different from the Ref.~\cite{c10}
.} :
\begin{equation}
 S_F = \sum_{x,y} \bar{\psi}(x) K(x,y) \psi(y),
\end{equation}
\begin{eqnarray}
 K(x,y) &=& \delta_{x,y}
     - \kappa_{\tau} \left\{ (1-\gamma_4)U_4(x)\delta_{x+\hat{4},y}
       + (1+\gamma_4)U_4^{\dag}(x-\hat{4})\delta_{x-\hat{4},y} \right\}
 \nonumber \\
 & & \hspace{0.7cm}
    - \kappa_{\sigma} \sum_{i}
      \left\{ (r-\gamma_i)U_i(x)\delta_{x+\hat{i},y}
      + (r+\gamma_i)U_i^{\dag}(x-\hat{i})\delta_{x-\hat{i},y} \right\}
 \nonumber \\
 & & \hspace{0.7cm}
    -  \kappa_{\sigma}  c_E \sum{i} \sigma_{4i}F_{4i}(x)\delta_{x,y}
    - r \kappa_{\sigma} c_B \sum_{ij} \frac{1}{2}
              \sigma_{ij}F_{ij}(x)\delta_{x,y}.
 \label{eq:action}
\end{eqnarray}
The spatial and the temporal hopping parameters, 
$\kappa_{\sigma}$ and  $\kappa_{\tau}$ respectively, are related to
the bare quark mass $m_0$ and
the bare anisotropy parameter $\gamma_F$ as follows.
\begin{eqnarray}
\kappa_{\sigma} &=& \frac{1}{2(m_0+\gamma_F+3r)}, \\
\kappa_{\tau} &=& \gamma_F \kappa_{\sigma}.
\end{eqnarray}
where the $m_0$ is in the spatial lattice unit.
In the free quark case, the bare anisotropy $\gamma_F$ is taken to be the
same value as the cutoff anisotropy $\xi=a_{\sigma}/a_{\tau}$.
In practical simulation, the anisotropy parameter receives the quantum
effect and should be tuned to give the same renormalized anisotropy
for the fermion and the gauge fields.
This ``calibration'' will be described later.

There have been used two choices of the value of the Wilson parameter $r$
for the anisotropic $O(a)$ improved quark action.
In this work, we adopt the choice\cite{c8}  $r=1/\xi$.
In this case, the temporal and the spatial directions are 
treated in the equal manner in the physical unit.
As the result, the tree level dispersion relation holds the
axis-interchange symmetry in the lowest order of $\vec{p}^2$.
On the other hand, this choice decreases the masses of doublers
which are introduced by the Wilson term to eliminate the unwanted
poles at the edges of the Brillouin zone. 
The dispersion relation is examined in the later part of this section.
Alternative choice, $r=1$, are adopted in the Ref.~\cite{c12,c13,c14}.
In this case, the contribution of doublers would not cause any problem,
in the cost of manifest axis-interchange symmetry.
Since we aim to develop the form applicable to the whole quark mass
region, $r=1/\xi^{-1}$ seems preferable especially in the light quark
mass region.

Here it is useful to define $\kappa$ so that which has
the same relation with the bare quark mass $m_0$ as the isotropic case:
\begin{eqnarray}
\frac{1}{\kappa} = \frac{1}{\kappa_{\sigma}} - 2(\gamma_F+3r-4)
 \hspace{0.5cm} (\, = 2(m_0+4) \, ). 
\end{eqnarray}
For the light quark systems, the extrapolation to the chiral limit
would be performed in $1/\kappa$.
The coefficients of the clover terms, $c_E$ and $c_B$, depend on the
Wilson parameter $r$.
In our choice $r=\xi^{-1}$,  $c_E$ and $c_B$ are unity at the tree level.

We apply the mean-field improvement proposed in the Ref.~\cite{c15}.
On the anisotropic lattice, the mean-filed values of the spatial 
link variable $u_{\sigma}$ and the temporal one $u_{\tau}$ are
different from each other.
The improvement is achieved by rescaling the link variable
as $U_i(x) \rightarrow U_i(x)/u_{\sigma}$ and  $U_4(x) \rightarrow
U_4(x)/u_{\tau}$.
This replacement leads the following values for the coefficient of
the clover terms.
\begin{equation}
 c_E=\frac{1}{u_{\sigma} u_{\tau}^2},~~~c_B=\frac{1}{u_{\sigma}^3}.
\label{eq:cecb}
\end{equation}
The determination of mean-field values of the link variable $u_{\sigma}$ 
and $u_{\tau}$ are described in the next section.

In this paper, our target mass region is around the charm quark mass.
The temporal cutoffs in this work are
$4.5$ and $6.4$ GeV, and well above the charm quark mass.
For these quark mass and $a_{\tau}^{-1}$, 
the effective-theoretical treatment
would not be necessary to be applied.
Such consideration will be called for the calculations containing
the $b$-quark on the same size of lattice.
In the effective-theoretical treatment, the ratio of the spatial
and the temporal hopping parameter is tuned so that they give
correct dispersion relation of the nonrelativistic quark\cite{c10}.
On the anisotropic lattice, the calibration automatically incorporates
this condition if one use the nonrelativistic dispersion relation
as the anisotropy condition.

\subsection{Dispersion relation of free quark}
\vspace*{-1pt}
\noindent
Now we consider how the dispersion relation of the free quark
is changed by the introduction of anisotropy.
Observing the action (\ref{eq:action}), one notices
that the larger anisotropy $\xi$ causes the smaller spatial Wilson term.
Then the question is how the contribution of the doubler eliminated
by the Wilson term becomes significant.
The action (\ref{eq:action}) leads the free quark propagator,
\begin{equation}
S(p) = \frac{1}{i\gamma_4 \sin{p_4}
     + i \; \zeta  \vec{\gamma}\cdot \vec{S}
     + m_t + (1-\cos p_4) + \frac{1}{2} r\zeta \vec{\hat{p}}^2},
\end{equation}
where $\zeta=\xi^{-1}$, $m_t=m_0\zeta$, $S_i=\sin p_i$ and 
$\hat{p_i}=2\sin(p_i/2)$.
Then the dispersion relation of the free quark is
\begin{equation}
\cosh E(\vec{p}) = 1 + \frac{\zeta^2 \vec{S}^2
            + (m_t + \frac{1}{2} r \zeta \vec{\hat{p}}^2 )^2}
            { 2 (1+m_t + \frac{1}{2} r\zeta \vec{\hat{p}}^2 ) }.
\label{eq:disp}
\end{equation}
Neglecting the higher order terms in $\vec{p}$ and in $a$,
the relativistic dispersion
relation $E^2=m^2 + \zeta^2 \vec{p}^2$ holds for the small quark mass.
($m^2=m_t^2(1-m_t)$, and $m_t$ is the bare parameter.)

\begin{figure}[tb]
\center{
\leavevmode\psfig{file=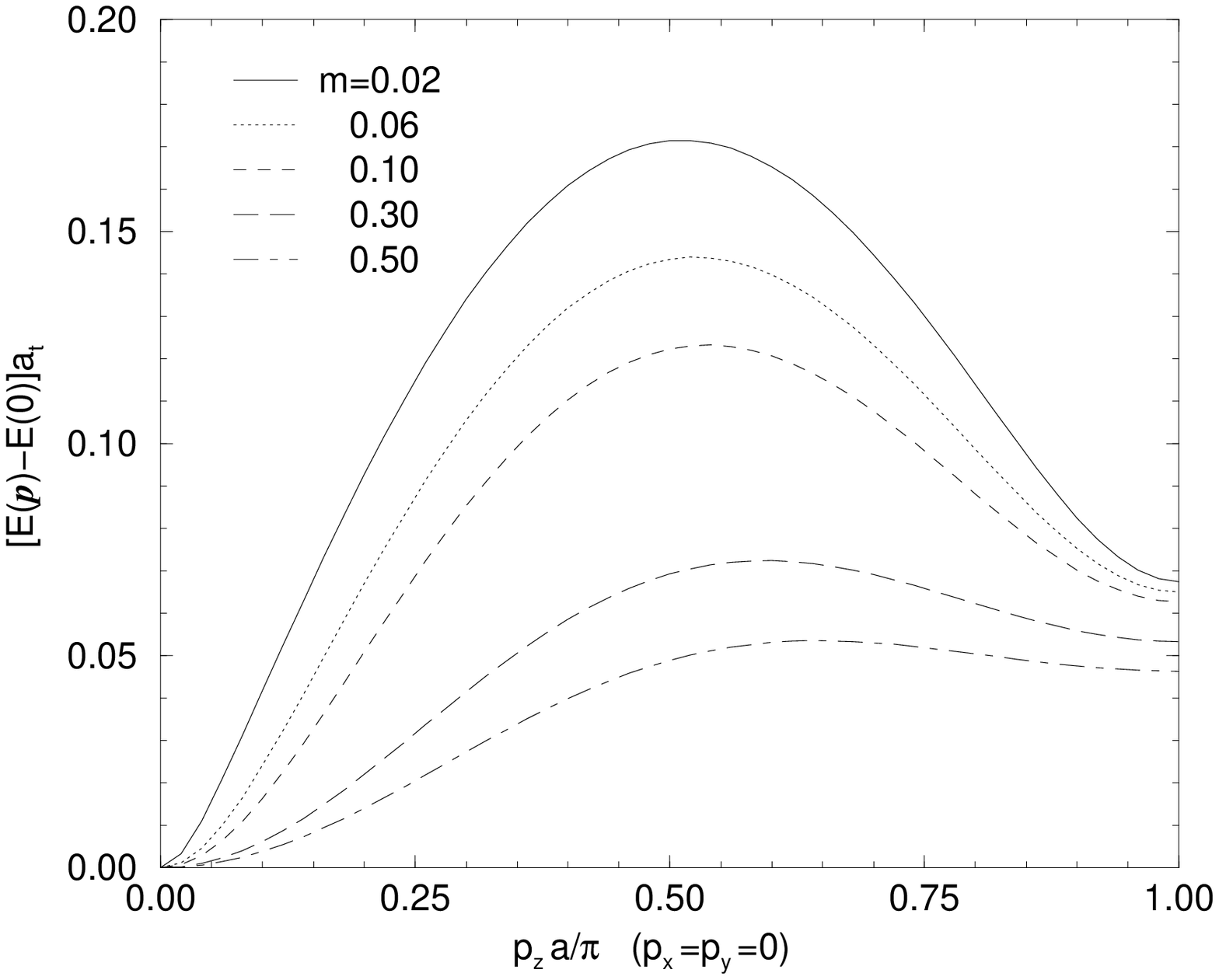,width=7.0cm}
\leavevmode\psfig{file=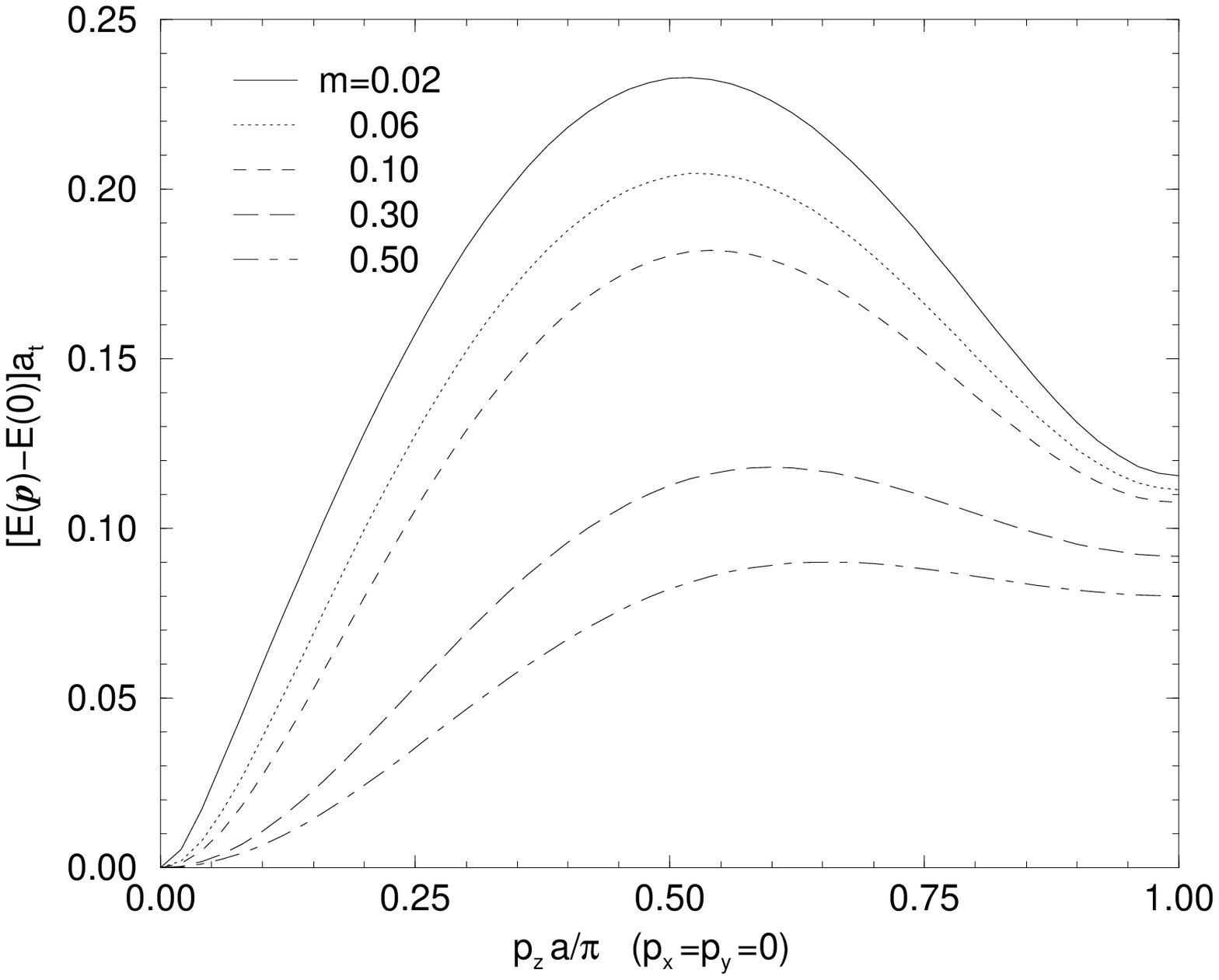,width=7.0cm}}
\caption{
Dispersion relation of anisotropic free quark.
Left figure is of Set-I, right one is
of Set-II.}
\label{fig:disp}
\end{figure}

Fig.~\ref{fig:disp} shows the dispersion relation (\ref{eq:disp})
at $\xi=5.3$ and $4.0$ for various values of $m_t$.
Now let us consider the practical cases that $a_{\tau}^{-1}=4.5$ GeV
for $\xi=5.3$ (Set-I)
and $a_{\tau}^{-1}=6.4$ GeV for $\xi=4.0$ (Set-II).
These values are obtained in our numerical simulations, and described
in the next section.
In the heavy quarkonium, the typical energy and momentum
exchanged inside the meson are in the order of $mv^2$ and $mv$ 
respectively\cite{c16}.
For the charmonium, $v^2 \sim 0.3$, then typical scale
of the kinetic energy is around $500$ MeV.
It is noted that
$m_t \sim 0.3$ for Set-I and $m_t \sim 0.2$ for Set-II correspond
to the charm quark mass.
Let us consider two quarks inside meson with opposite momenta $p=\pm \pi/a$.
Then $2(E(p_z=a/\pi)-E(0)) \sim 0.5$ GeV and $\sim 1$ GeV for
Set-I and Set-II respectively.
Although Set-I lattice may not be free from the systematic effect,
Set-II would be successfully applicable to the low-lying charmonium
system.

For comparison, we also examine the light quark mass region. 
For Set-I, $m=0.02$--$0.06$ corresponds to 90-270 MeV,
which is used in the Ref.~\cite{c8}
as the light quark mass region with the
anisotropic Wilson quark action with $r=1$.
$E(\vec{p})-E(0)$ rapidly decrease at the edge of the Brillouin zone,
and the height at $z=a/\pi$ is around 300 MeV.
For two quarks with momenta $p=\pm a/\pi$, additional energy of
doublers is $\sim 600$ MeV,
and again seems not sufficiently large compared with the typical
energy scale transfered inside mesons.
In the case of Set-II, this value increases to 1.4 GeV, and seems to
be applicable to the meson systems without large systematic effect.

\subsection{Calibration of quark field}
\vspace*{-1pt}
\noindent
On anisotropic lattice, the anisotropy of quark configuration $\xi_F$
must be equal to that of  gauge field $\xi$.
\begin{equation}
 \xi(\gamma_G,\gamma_F)= \xi_F(\gamma_G,\gamma_F)
\label{eq:xi}
\end{equation} 
Since $\xi$ and $\xi_F$ are function of $\gamma_G$ and $\gamma_F$ in
general, 
a nonperturbative determination of the combination of $\gamma_G$ and
$\gamma_F$ which satisfy the condition (\ref{eq:xi}) requires much
effort.
In the quenched case, however, these determination are rather easy to
be performed, because $\xi$ can be determined independently of $\gamma_F$.
After the determination of $\xi$, one can tune
$\gamma_F$ so that the certain observable satisfies the condition
(\ref{eq:xi}).
We call this procedure as ``calibration''.

There are several determinations of $\xi_F$.
In the Ref.~\cite{c8},
 the ratio of the temporal to the spatial meson masses,
$\xi_F=m_\sigma/m_\tau$, is used as such a observable.
However this is not suitable for the present case, 
since the charm quark mass is larger than or comparable with
the spatial lattice cutoffs.
In order to determine $\xi_F$, we use the dispersion relation of
the free meson\cite{c12,c13,c14}.
For a heavy quark, one may use the nonrelativistic dispersion relation,
$E=m +p^2/2m \xi_F^2$.
In this paper, we alternatively use the relativistic dispersion relation
of meson for the calibration. 
This form is also available for the light quark mass region.

We assume that the meson is described by the following
lattice Klein-Gordon action.
\begin{equation}
S = \sum_{x} \frac{1}{2\xi_F} \phi^{\dag}(x) \left[
     - \xi_F^2 D_4^{\,2} - \vec{D}^2 + m_0^2  \right] \phi(x),
\end{equation}
where $m_0$ is in the unit of $a_{\sigma}$.
Then the free meson satisfies the dispersion relation
\begin{equation}
\cosh E(\vec{p}) = 1 + \frac{1}{2\xi_F^2} (\vec{\hat{p}}^2+m_0^2).
\label{eq:disp2}
\end{equation}
Using this relation, one can determine the anisotropy $\xi_F$ as
\begin{equation}
\xi_F^2
 = \frac{\vec{\hat{p}}^2}{ 2( \cosh E(\vec{p}) - \cosh E(0) ) }
\label{eq:disp3}
\end{equation}
This condition forces the axis-interchange symmetry to the
meson field.

\section{Lattice setup}
\label{sec:setup}
\subsection{Gauge configuration}
The numerical calculations performed on the two sets of anisotropic
lattices\cite{c17}, 
one is with the standard plaquette action and the other is with Symanzik
type improved action at the tree level\cite{c18}.
These actions are represented with the following form;  
\begin{eqnarray}
 S_G &=& \frac{\beta}{\gamma_G}\sum_{x,i<j\leq 3} \left[ c_{11}
(1-P_{ij}(x)) + c_{12}(2-R_{ij}(x)-R_{ji}(x))\right] \nonumber\\
     &&+ \beta \gamma_G\sum_{x,i\leq 3}\left[ c_{11}
(1-P_{i4}(x)) + c_{12}(2-R_{i4}(x)-R_{4i}(x)) \right],
\end{eqnarray}
where the plaquette $P_{\mu \nu}(x)$
and rectangular loop $R_{\mu \nu}(x)$ are defined as follows,
\begin{equation}
P_{\mu \nu} \equiv \frac{1}{3}\mbox{Re}\mbox{Tr}
[U_\mu(x)U_\nu(x+\hat{\mu})U^\dag_\mu(x+\hat{\nu})U^\dag_\nu(x)] 
\end{equation}
\begin{equation}
R_{\mu \nu} \equiv \frac{1}{3} \mbox{Re}\mbox{Tr}
[U_\mu(x)U_\mu(x+\hat{\mu})U_\nu(x+2\hat{\mu})
U^\dag_\mu(x+\hat{\nu}+\hat{\mu})U^\dag_\mu(x+\hat{\nu})
U^\dag_\nu(x)]. 
\end{equation}

The standard action is the case with $c_{11}=1$ and $c_{12}=0$, 
and the improved action is $c_{11}=5/3$ and $c_{12}=-1/12$. 
The Set-I is the same configurations as used in the Ref.~\cite{c8}.
The parameters for these configurations are summarized in
Table~\ref{tab:input}.
These parameters are adopted so that the spatial and temporal lattice
extent are about 3 fm respectively.

Both numerical calculations are done on the quenched configurations
which are generated by the pseudo-heat bath algorithm with
20,000 thermalization sweeps, the configurations being separated by
2,000 sweeps. These configurations are fixed to Coulomb gauge.
The statistical errors are estimated using the jackknife method unless
mentioned explicitly.

\begin{table}[tb]
\begin{center}
\begin{tabular}{ccccccc}
\hline \hline
 Set & $c_{11}$ & $c_{12}$ & size & $\beta$ & $\gamma_G $ & \# conf. \\
\hline
Set-I & 1 & 0 & $12^3\times N_t$ & 5.68 & 4.00 & 60 \\
Set-II& 5/3 & -1/12 & $16^2\times 24\times N_t$ & 4.56 & 3.45 & 120 \\
\hline \hline
\end{tabular}
\end{center}
\caption{Simulation parameters for the gauge configurations.}
\label{tab:input}
\end{table}

We determine the parameters for the gauge field configurations, renormalized
anisotropy $ \xi (\equiv a_{\sigma}/a_{\tau} ) $ and spatial cutoff scale 
$ a^{-1}_{\sigma} $. 
We define the ratios of the Wilson loop on spatial-spatial 
($\sigma \sigma$) and spatial-temporal ($\sigma \tau$) 
plane\cite{c19,c20,c21}.
\begin{eqnarray}
 R_{\sigma}(r,x)&=&W_{\sigma \sigma}(r,x)/W_{\sigma \sigma}(r+1,x),\\
 R_{\tau}(r,t)&=&W_{\sigma \tau}(r,t)/W_{\sigma \tau}(r+1,t)
\end{eqnarray}
Then $\xi$ is determined so that the matching condition, 
$R_{\sigma}(r,x)=R_{\tau}(r,t=\xi x)$, is satisfied.
From this analysis we conclude $\xi=5.3(1)$ (Set-I) and 
$\xi=3.950(18)$  (Set-II).
Cutoff scales are determined from the static quark potential
using the physical value of string tension 
$\sqrt{\sigma_{\mbox{\tiny phys}}}= 427 \mbox{MeV}$\cite{c22}.
The spatial cutoff scales are $a_{\sigma}^{-1} =0.85(3)$ GeV (Set-I) and 
$1.610(14)$ GeV (Set-II) respectively. 

From the calculation of Polyakov loops and its susceptibilities,
we find that the temperatures of lattices with $N_t=18$
(Set-I) and $24$ (Set-II) are both just above $T_c$.
Then we estimate $T_c=250 \sim 260$ MeV from the temporal cutoff scale
on our lattices. These estimation are consistent with the
other quenched lattice results.
We prepare the finite temperature gauge configurations at just below
$T_c$, just above $T_c$ and about $1.5T_c$ for each sets.
These results are summarized in Table~\ref{tab:output}. 

\begin{table}[tb]
\begin{center}
\begin{tabular}{clllc}
\hline \hline
Set & ~~$\xi$ & $a_{\sigma}^{-1}$(GeV) & $a_{\tau}^{-1}$(GeV) &
 $N_t(T/T_c)$  \\ 
\hline
Set-I & 5.3(1) & 0.85(3) & 4.5(2) 
& 72~($\sim$ 0),~20~(0.93),~16~(1.15),~12~(1.5)~  \\
Set-II & 3.950(18) & 1.610(14) & 6.359(62) 
& 96~($\sim$ 0),~26~(0.93),~22~(1.10),~16~(1.52)  \\
\hline \hline
\end{tabular}
\end{center}
\caption{Scale parameters for the gauge configuration. The $\xi_F$,
 $a_{\sigma}^{-1}$ and $a_{\tau}^{-1}$ are measured on the
 configurations at $T=0$ ( $N_t=72~(\mbox{Set-I}),~96~(\mbox{Set-II})$).
Here the temperatures are roughly estimated in the unit of $T_c$.}
\label{tab:output}
\end{table}

Here we notice that the hadronic correlator depends on which the sector
Polyakov loop stays in $Z_3$ space at $T>T_c$. 
Since we treat the quenched QCD as an approximation of the full QCD, we 
choose the real sector for the value of Polyakov loop at $T>T_c$.

\subsection{Mean-field value on the anisotropic lattice}
\label{sec:MF}
\vspace*{-0.1pt}
\noindent
We state the calculation of mean-filed value on the anisotropic lattice 
for the mean-field improvement of the clover coefficients.

There are two commonly used methods.
One is determined from the expectation value of plaquette. This is widely
used for its easiness to measure. On anisotropic lattice these are
determined as follows.
\begin{equation} 
 u_{\sigma}=\frac{1}{3}\langle \mbox{Re}\mbox{Tr}
P_{ij}(x)\rangle^{\frac{1}{4}}, 
~~u_{\tau}=\frac{1}{3 u_{\sigma}}\langle\mbox{Re}\mbox{Tr}
P_{i4}(x)\rangle^{\frac{1}{2}}
\end{equation} 
 The other is the trace of link calculated in the Landau gauge. 
\begin{equation} 
 u_{\sigma} = \frac{1}{3}\langle \mbox{Re}\mbox{Tr}U_{i}(x)\rangle,
~~u_{\tau} = \frac{1}{3}\langle \mbox{Re}\mbox{Tr}U_{4}(x)\rangle
\end{equation}
In the former case $u_{\tau}$ is greater than the unity on our lattice. 
Then the latter definition seems more reasonable.
In this work we determined the mean-field values in the Landau gauge.

The Landau gauge fixing is realized by maximization of
Eq.(\ref{eq:landau}). 
\begin{equation}
\sum_x \mbox{Re} \mbox{Tr} 
\left( U_1(x)+U_2(x)+U_3(x)+c_{st} U_4(x)\right) 
\label{eq:landau}
\end{equation}
Here, the temporal coefficient $c_{st}$ appears on the
anisotropic lattice.  
We adopt the self-consistent mean-field improvement of $c_{st}$.
As the tree level we chose $c_{st}=\xi^2$.
Using the mean-field improved $c_{st}=\xi^2 u_{\sigma}/u_{\tau}$  
we calculate the mean-field values recursively.
We perform these calculation with 20 configurations. 
In our case, the result of 3rd measurement is consistent with the input
$u_{\sigma}/u_{\tau}$ which is determined from the linear interpolation of
the tree level and first mean-field improvement result.
These results are summarized in Table~\ref{tab:MF} together with the
mean-field improved $c_E$ and $c_B$. 
Here we mention that the mean-field improved $\gamma_G$
give the reasonable estimation for $\xi$ within 1\% (Set-I) and 6\%
(Set-II) error. 
\begin{table}[tb]
\begin{center}
\begin{tabular}{cllcc}
\hline \hline
 Set & $u_{\sigma}$ & $u_{\tau}$ & $c_E$ & $c_B$ \\ 
\hline
Set-I & 0.75050(16) & 0.992436(13) & 1.7889 & 2.3656 \\ 
Set-II& 0.812354(92) & 0.9900788(90) & 1.5305 & 1.8654 \\ 
\hline \hline
\end{tabular}
\end{center}
\caption{Mean-field value of link variable and mean-field improved
 clover coefficients for each sets. }
\label{tab:MF}
\end{table}

\subsection{Calibration result}
\label{sec:calib}
\begin{figure}[tb]
\center{
\leavevmode\psfig{file=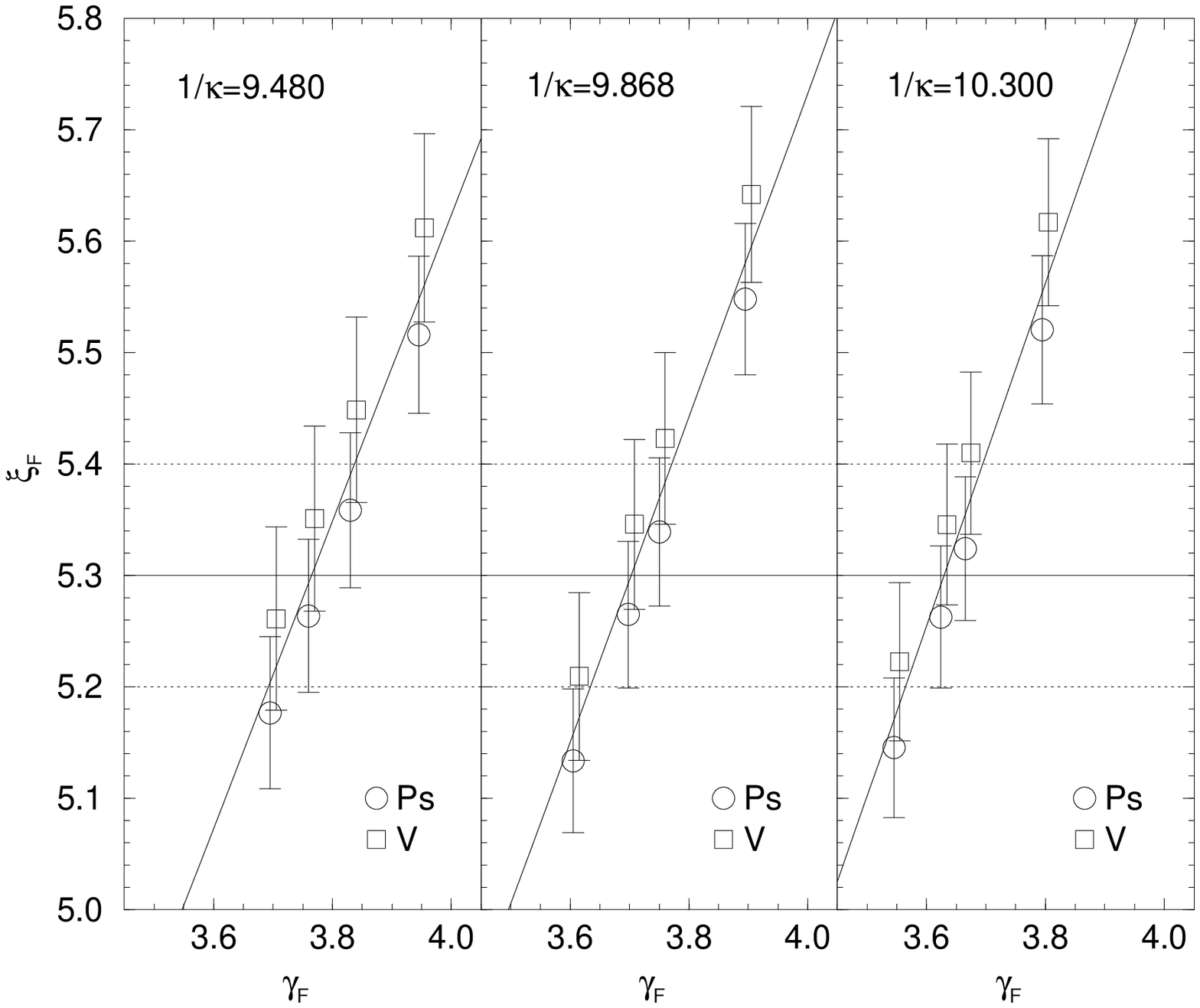,width=70mm}
\leavevmode\psfig{file=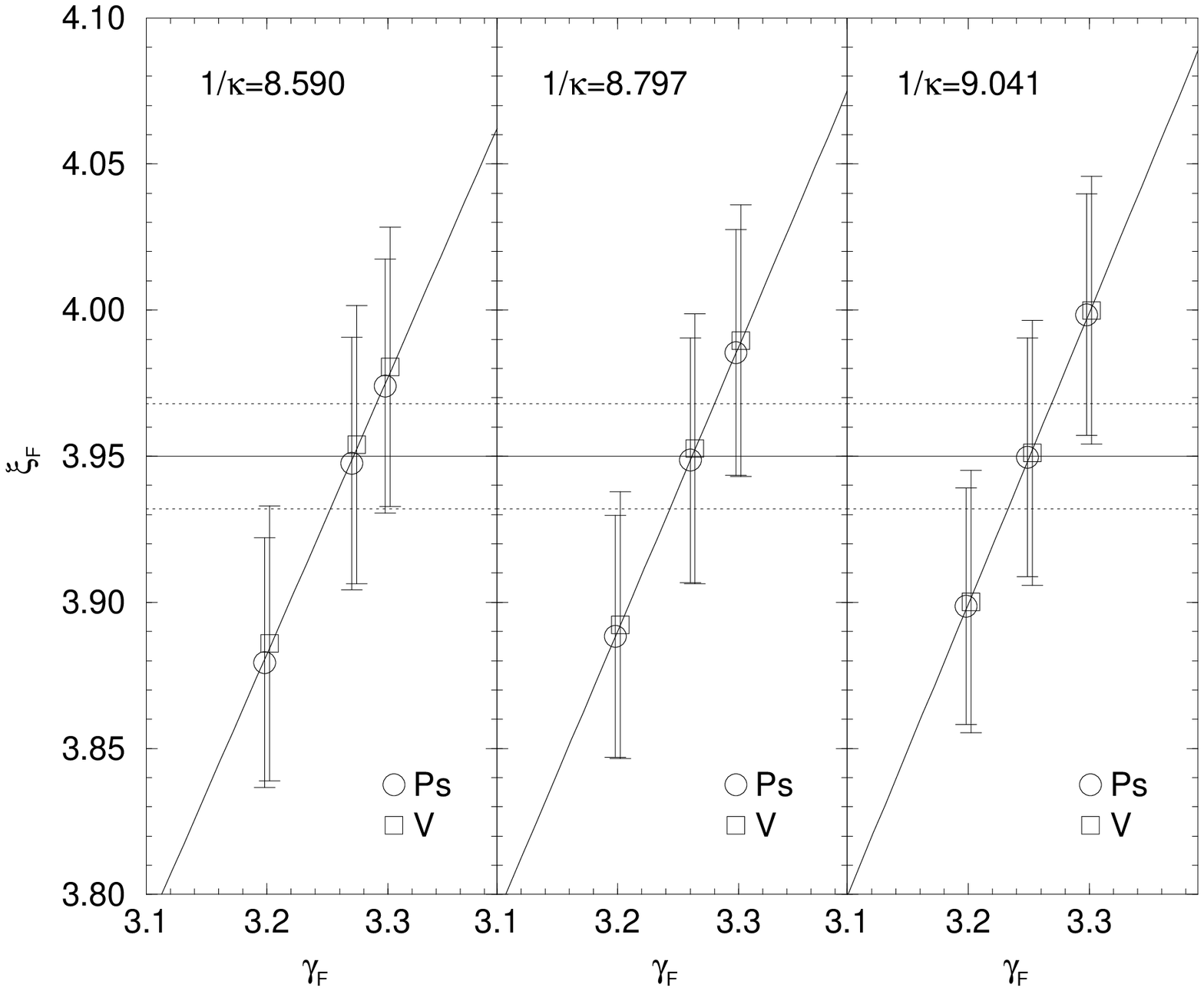,width=70mm}}
\caption{The left figure shows the calibration for Set-I and the right
 one is for Set-II. 
In the both figures, values of $\xi$ ( =5.3(1) (Set-I) and =3.950(18)
 (Set-II)) are indicated.
The three values of $\kappa$ correspond to a pseudoscalar mass 
$m_{\mbox{\tiny Ps}}\simeq 2.6,~3.0$ and $3.4$ GeV respectively.} 
\label{fig:calib}
\end{figure}

\vspace*{-0.5pt}
\noindent
We now turn to the calibration of quark field.
It is performed by the Eq.(\ref{eq:disp2}) with the momentum 
$\vec{p}=(0,0,0)$ and $(0,0,1)$ for each mesonic channel, pseudoscalar
(Ps) and vector (V).  
These results are summarized in Fig.~\ref{fig:calib}.
These calibrations are performed at the parameters which correspond to the 
pseudoscalar mass $m_{Ps}\simeq 2.6,~3.0$ and $3.4$ GeV respectively.

The detailed definition of meson correlator which is adopted here is
shown in Sect.~4.1 
In these calculations we use the smeared source function which is
determined from the wave function at $1/\kappa=9.868$ (Set-I) and
$\kappa=8.797$ (Set-II).
Here we mention that the shape of wave function has very weak $\kappa$ 
dependence.

From the Fig.~\ref{fig:calib}, we find that the $\gamma_F$ dependence of
$\xi_F$ is almost linear and $\kappa$ dependence of its slope is small.
The $\gamma_F$ which is determined from the pseudoscalar and
the vector gives consistent value within the statistical error for each
$\kappa$. 
Especially $\xi_F$ for Set-II are in good agreement with the
pseudoscalar and the vector.
We determine the $\gamma_F$ which satisfies the condition of
Eq.(\ref{eq:xi}) from the interpolation with the linear fitting.
The error of $\gamma_F$ is estimated from the error of $\xi$ and $\xi_F$.
These calibration results are summarized in Table~\ref{tab:calib}.
Here we mention that the mean-field improved $\gamma_F$
give the reasonable estimation for $\xi_F$ within 9\% (Set-I) and 1\%
(Set-II) error.

\begin{table}[tb]
\begin{center}
\begin{tabular}{cccc}
\hline \hline
 & $1/\kappa$ & $\kappa_s$ & $\gamma_F$ \\ 
\hline 
      & 10.300 & 0.093545 & 3.629(91) \\ 
Set-I &  9.868 & 0.096098 & 3.703(84) \\ 
      &  9.480 & 0.098599 & 3.765(79) \\ 
\hline
      & 9.041 & 0.110350 & 3.251(52) \\ 
Set-II& 8.797 & 0.113121 & 3.262(49) \\ 
      & 8.590 & 0.115564 & 3.272(47) \\ 
\hline \hline
\end{tabular}
\end{center}
\caption{Calibration results for Set-I and Set-II.
These parameters satisfy the condition of $\xi=\xi_F$ within the
statistical error for the both mesonic channel. The error of
 $\gamma_F$ is estimated from the error of $\xi$ and $\xi_F$ }
\label{tab:calib}
\end{table}

\section{Results at Zero temperature}
\label{sec:zero}
\subsection{Meson correlators}
\label{sec:corr}
The mesonic operator which is used in this paper is in the following
form,
\begin{equation}
 {\cal O}^{(\omega)}_M(\vec{z},t) = \sum_{\vec{y}} 
\omega(\vec{y})\bar{q}(\vec{z},t)\gamma_M q(\vec{z}+\vec{y},t),
\label{eq:smear}
\end{equation}
where $\omega(\vec{y})$ is a smearing function and 
$\gamma_M= \gamma_5$ and $\gamma_1$ for 
the pseudoscalar and the vector respectively.
Using these mesonic operators we construct the meson correlators
$G_M^{\omega' \leftarrow \omega}(t)$ in the Coulomb gauge
as the following,
\begin{eqnarray}
 G_M^{\omega' \leftarrow \omega}(t)&=&\sum_{\vec{z}}\langle 
{\cal O}^{(\omega')}_M(\vec{z},t) {\cal O}^{(\omega)\dagger}_M(0)
\rangle
\end{eqnarray}
According to our strategy, we need to
construct a good operator which has large overlap with the ground
state ( or the excited state ) of heavy quarkonium.
For this purpose we examine two types of correlators.

In the first type of correlator we use the smeared source 
and point sink  
( $\omega'(\vec{y})\propto \delta(\vec{y})$ ), which is 
already used in the calibration described in Sect.~3.3 .
This source function is defined from the measured wave function
$w(\vec{x})$\cite{c23} so that the smearing reflects the actual
distribution of quark and anti-quark. 
This smearing procedure works well for the suppressing higher excited
state to the correlator.
The wave function is well fitted to the form
\begin{equation}
w(\vec{y})\propto \exp{(-a|\vec{y}|^p)}.\label{eq:exp}
\end{equation} 
The determined parameters $a$ and $p$ are used to generate the smearing
function. 
These correlators are used in the charmonium spectroscopy in the next
subsection.

For the study at $T>0$ we need more systematic optimization of
the correlators.
We apply the variational analysis, and regard the diagonalized correlator
as the optimized correlator.
This analysis can extract not only the ground state but also the excited
state. 
In this analysis we use the correlators with  smeared source and sink,
whose smearing functions are determined by solving the Schr{\" o}dinger
equation with the potential model. 
This technique is mentioned in Sect.~4.3 .

\subsection{Charmonium spectroscopy}
\begin{figure}[tb]
\center{
\leavevmode\psfig{figure=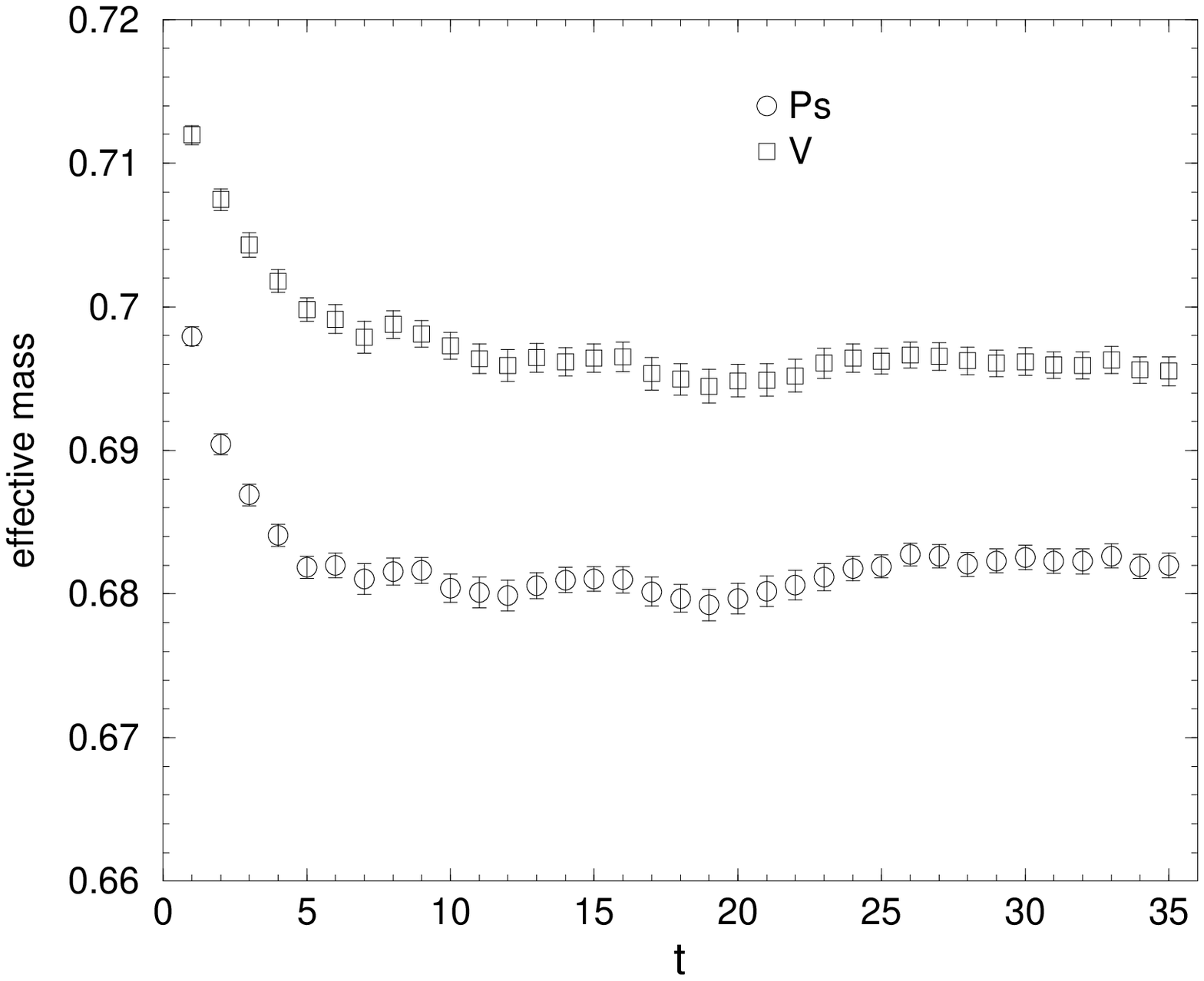,height=2.2in}
\leavevmode\psfig{figure=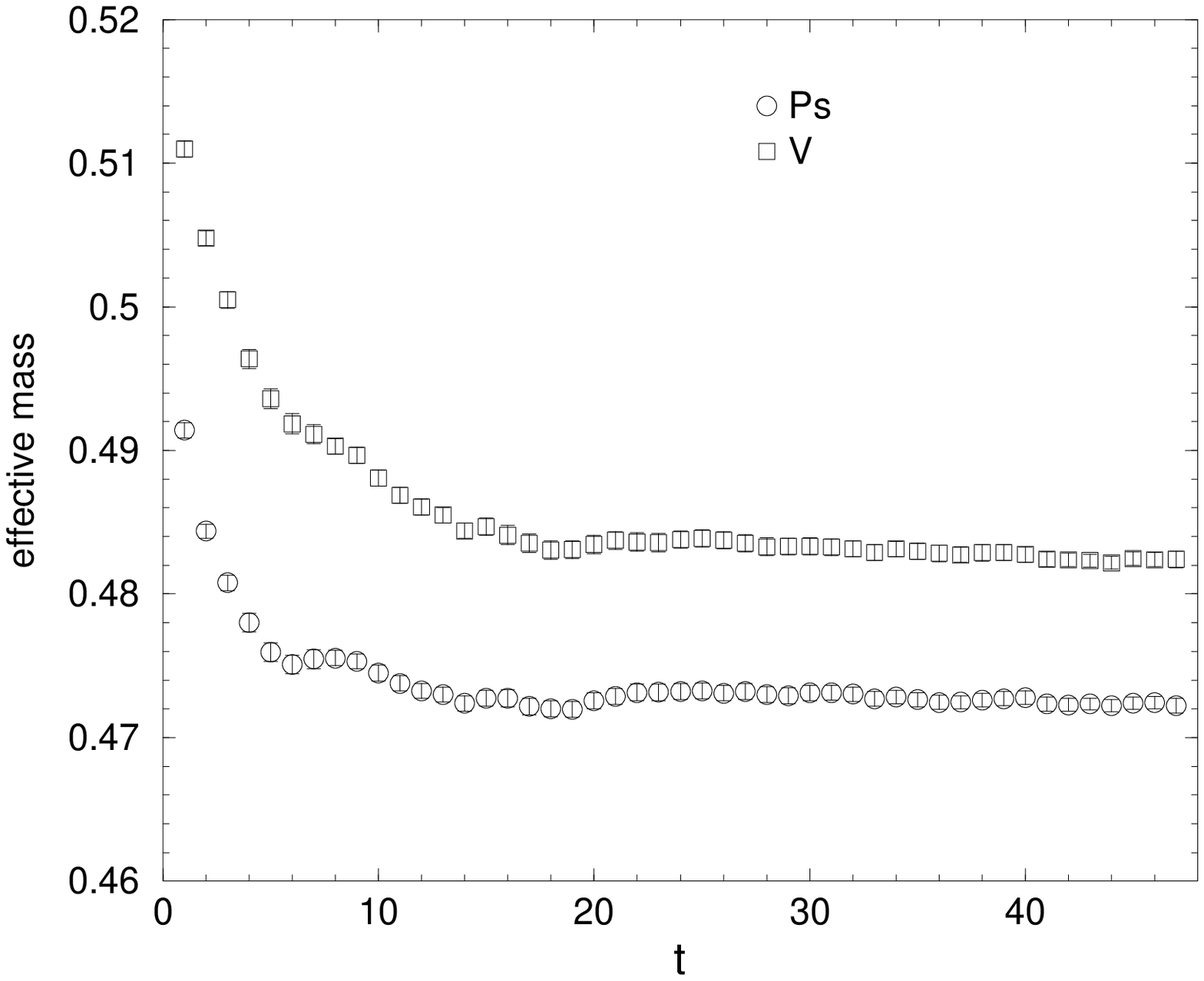,height=2.2in}}
\caption{Effective masses at $T=0$. The left figure is for Set-I
and the right one is Set-II.}
\label{fig:eff0}
\end{figure}

\vspace*{-0.5pt}
\noindent
We show the results of spectroscopy for the heavy quarkonium at $T=0$,
which is the basis of the study at $T>0$.
In this section we also examine how our action works in the heavy
quark system. 
The pseudoscalar and the vector meson channels are measured with the
parameters listed in Table~\ref{tab:calib}. 
Figure~\ref{fig:eff0} show the effective mass plot at $T=0$. 
In this calculation we use the correlators with the smeared source
introduced in the previous subsection.  
Here the parameters for the smearing function are 
$a=0.7096,~p=1.1267$ for Set-I and $a=0.2978,~p=1.1918$ for Set-II.
The plateaus of the effective mass appear beyond $t\simeq 10$ (Set-I), 
$20$ (Set-II).
Concerning the determination of smearing function our choice
works well in heavy quark system. 
On the other hand in the light quark system the wider smearing function 
enhances the ground state contribution\cite{c8}. 
In the small $t$ region, contributions from the excited states remain.
However this small $t$ region is the main stage of the study at $T>0$.
In order to estimate the temperature effects of the ground state
contribution at the high temperature,   
we need further improvement of the mesonic operator.
In the next subsection more systematic study with variational analysis
is examined.

We fit the correlator to the single exponential form at $t=30$--$36$
(Set-I), $40$--$48$ (Set-II).
These results are summarized in Table~\ref{tab:spec}. 
\begin{table}[tb]
\begin{center}
\begin{tabular}{ccccc}
\hline \hline
 & $1/\kappa$ & $m_{\mbox{\tiny Ps}}$ & $m_{\mbox{\tiny V}}$ & HFS \\ 
\hline 
      & 10.300 & 0.77934(75) & 0.79146(83) & 0.01213(26)\\
Set-I &  9.868 & 0.68235(77) & 0.69600(88) & 0.01365(30)\\
      &  9.480 & 0.59381(78) & 0.60917(91) & 0.01535(35)\\
\hline
      & 9.041 & 0.54170(39) & 0.55050(43) & 0.00880(15)\\
Set-II& 8.797 & 0.47236(39) & 0.48238(44) & 0.01002(16)\\
      & 8.590 & 0.41029(40) & 0.42182(45) & 0.01152(18)\\
\hline \hline
\end{tabular}
\end{center}
\caption{Spectroscopy for Set-I and Set-II with parameters in
 Table~\ref{tab:calib}. HFS means the hyperfine splitting for a heavy
 quarkonium, $m_{\mbox{\tiny Ps}}-m_{\mbox{\tiny V}}$. There results are
 shown in $a_{\tau}$ unit.} 
\label{tab:spec}
\end{table}
The parameters with $1/\kappa=9.868$ (Set-I)
and $1/\kappa=8.797$ (Set-II) in Table~\ref{tab:spec}
correspond approximately to the charm quark. Therefore we study the meson
correlators with these parameters in the successive calculations.

\begin{figure}[tb]
\begin{center}
 \leavevmode\psfig{figure=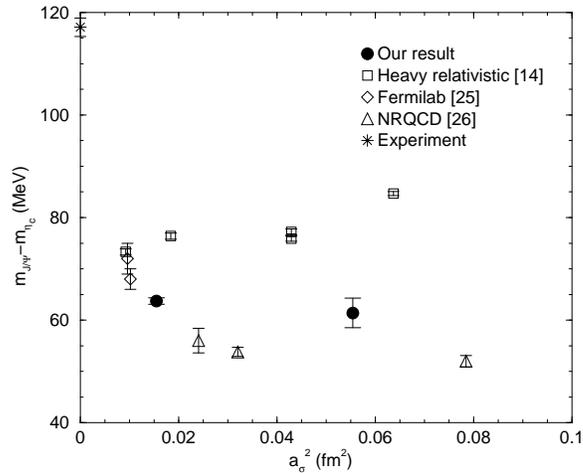,height=2.5in}
\end{center}
\caption{Hyperfine splitting for charmonium. On this figure results of
the other group are also shown. These groups\cite{c14,c25,c26} 
used the Sommer scale, the string tension and 1P-1S splitting for the
determination of the lattice cutoff scale respectively.}
\label{fig:hfs}
\end{figure}
Figure~\ref{fig:hfs} shows the hyperfine splitting for charmonium 
($m_{J/\psi}-m_{\eta_c}$).
For comparison, results of other group are shown simultaneously.
Here we notice that these results largely depend on how to
determine the lattice cutoff scale. 
Results of Fermilab action on the isotropic lattice\cite{c24}, whose
scale is determined from the physical value of the string tension, 
are roughly consistent with our results.  
Results of heavy relativistic action on anisotropic
lattice\cite{c13} and NRQCD\cite{c25} show the similar tendency. 
These scales are determined from the Sommer scale and the 1P-1S splitting
respectively. 
However, all the quenched results are roughly a half of experimental
results $117(2)$MeV\cite{c3}.
The differences from the experimental value of the hyperfine splitting
can be partly explained with the dynamical quark effects\cite{c26}.

\subsection{Variational analysis}
\label{sec:va}
\vspace*{-0.5pt}
\noindent
We examine the variational analysis for constructing more optimized 
operators. 
With this analysis we can construct series of operators which
have better matching to the states of interest.

Here we explain the principle of variational analysis briefly.
Firstly, we suppose that the state generated by mesonic operator on the
lattice is the linear combination of eigenstates for Hamiltonian.
Practically we assume that the generated states consist of $N$ linearly
independent states.  Then we prepare $N$ mesonic operators, which have
the same quantum numbers.
These operators are constructed with different smearing function
$w(\vec{x})$ in Eq.(\ref{eq:smear}). 
Then we get the $N\times N$ correlator matrix, $C_M^{ij}(t)$, 
as the following, 
\begin{equation}
C_M^{ij}(t) = \sum_{\vec{z}}
\langle {\cal O}_M^{(\omega_i)}(\vec{z},t)
{{\cal O}_M^{(\omega_j)}}^\dag (0) \rangle .
\label{eq:corr_m}
\end{equation}
Because of the symmetric nature of this matrix, we can get the
diagonalized correlators which are optimized correlators for the state of
interest.

\begin{figure}[tb]
\center{
\leavevmode\psfig{figure=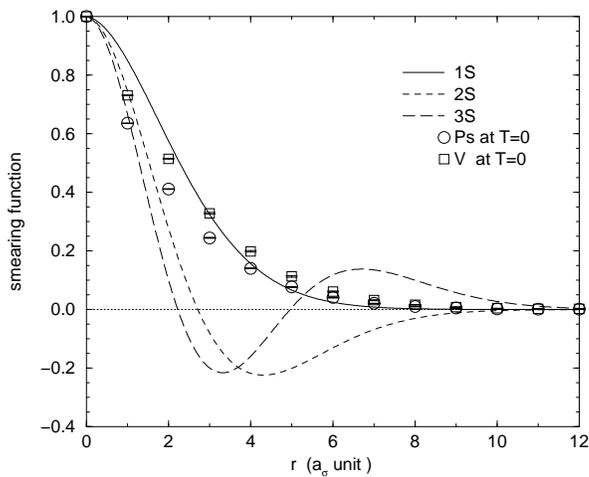,height=2.5in}}
\caption{Smearing functions for the variational analysis for Set-II.
These are calculated with the Schr{\" o}dinger equation with the potential 
model. The circle and square symbols are the wave function measured on the
lattice at $T=0$ for the pseudoscalar and the vector respectively.}
\label{fig:smear}
\end{figure}

In our study this variational analysis is applied in the minimal space
including the ground, such as 1st excited and 2nd excited state.  
We have to prepare the smearing functions so that this analysis works
well in this condition.
Thus we construct the smearing functions $\phi_l(r)$ using the 
Schr{\" o}dinger equation with the potential model,
\begin{eqnarray}
  \left[ -\frac{1}{2m_R}\frac{d^2}{dr^2}+\frac{l(l+1)}{2m_Rr^2}
 +V(r) \right]y_l(r) &=& E y_l(r),\nonumber\\
 y_l(r)&=&r\phi_l(r),
\end{eqnarray}
where $V(r)$ is the static quark potential measured on our lattice
and $m_R=1.5/2$ GeV.
The spin interaction is neglected and we calculate only S-state ($l=0$). 
Figure~\ref{fig:smear} shows the lowest three solutions $\phi_0(r)$
and the measured wave function at $T=0$ for Set-II.

\begin{figure}[tb]
 \begin{center}
  \leavevmode\psfig{figure=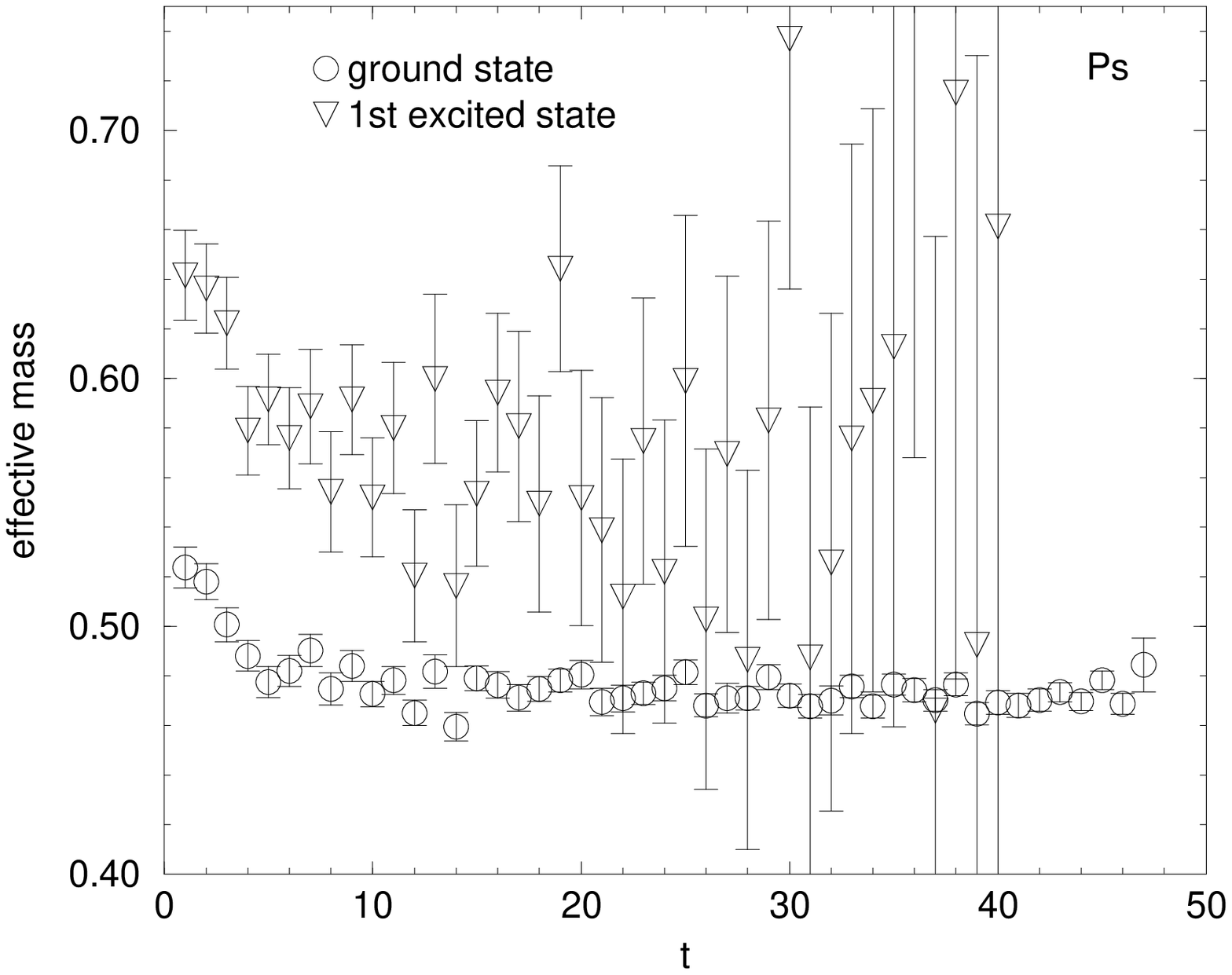,height=2.2in}
  \leavevmode\psfig{figure=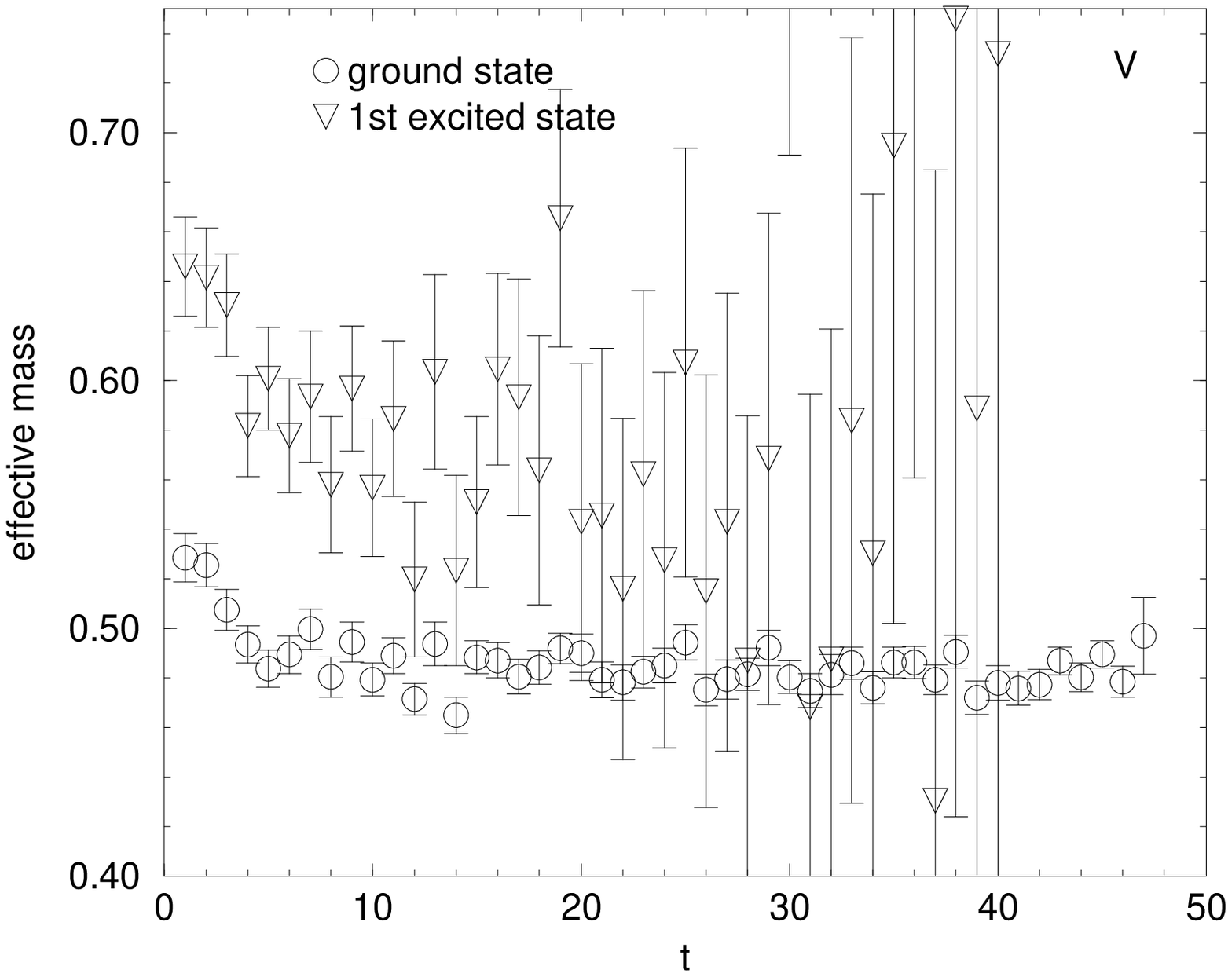,height=2.2in}
 \end{center}
 \caption{Effective mass of diagonalized correlator for Set-II at $T=0$.
The left figure is the result of pseudoscalar meson and the right one is
that of vector meson. Only ground and 1st excited one are shown because of
the large noise for the 2nd excited state. }
 \label{fig:va0}
\end{figure}

We calculate the diagonalized correlator using the three types of smearing
function in Fig.~\ref{fig:smear}.
The orthogonal matrices are obtained at each $t$, then
we adopt the averaged one as the orthogonal matrix which is
used for the calculation of the diagonalized correlator. 
With the orthogonal matrix we obtain the diagonalized correlator. Then its
effective mass plots (Set-II) are shown in Fig.~\ref{fig:va0}.
We fit the these data at $t=10$--$36$ for Set-I and $t=14$--$48$ for
Set-II, and get the results summarized in Table~\ref{tab:excited}.
\begin{table}[tb]
\begin{center}
\begin{tabular}{ccccc}
\hline \hline
 & Ps(2S) & V(2S) & $m_{\mbox{\tiny Ps(2S)}}-m_{\mbox{\tiny Ps(1S)}}$ 
 & $m_{\mbox{\tiny V(2S)}}-m_{\mbox{\tiny V(1S)}}$ \\ 
\hline 
Set-I & 0.7688(34) & 0.7794(38) & 0.0868(30) & 0.0848(33)\\
 ~~~(MeV) & 3460(155) & 3507(157)  &  391(22)   &  382(23)  \\
\hline
Set-II& 0.5641(66) & 0.5696(73) & 0.0913(64) & 0.0869(70) \\
 ~~~(MeV)     & 3587(55)   &  3622(58)  &   581(41)  &  553(45)  \\
\hline
Exp. value (MeV) \cite{c3} & 3594(5)& 3685.96(9)& 614(5) & 589.07(13)\\ 
\hline \hline
\end{tabular}
\end{center}
\caption{Spectroscopy for 2S state and 2S-1S splitting of charmonium.
For each sets the upper values are in $a_{\tau}$ unit and the lower one
 are in physical unit. The error in physical unit includes the error of
 $a_{\tau}$. } 
\label{tab:excited}
\end{table}

The extracted mass of 1S state are consistent with previous results in
Table~\ref{tab:spec}. 
In the Table~\ref{tab:excited} results are presented in
$a_{\tau}$ unit and physical unit for each sets, where the latter case
includes the error of $a_{\tau}$. 

The results of 2S-1S splitting are consistent with the experimental
values within a statistical error.
This is in contrast to the case of the hyperfine splitting in which we
obtain the half of the experimental value,
2S-1S splitting is in good agreement with experimental value. 

The variational analysis can directly extract one of correlators for
the ground and excited states.  
Therefore this analysis is useful to investigate the excited state
of hadron.
In the next section we mainly use this analysis at $T>0$.

\section{Results at Finite temperature}
\label{sec:ft}
\subsection{Temperature dependence of correlators}
\label{sec:tdep_corr}
\begin{figure}[tb]
\center{
\leavevmode\psfig{figure=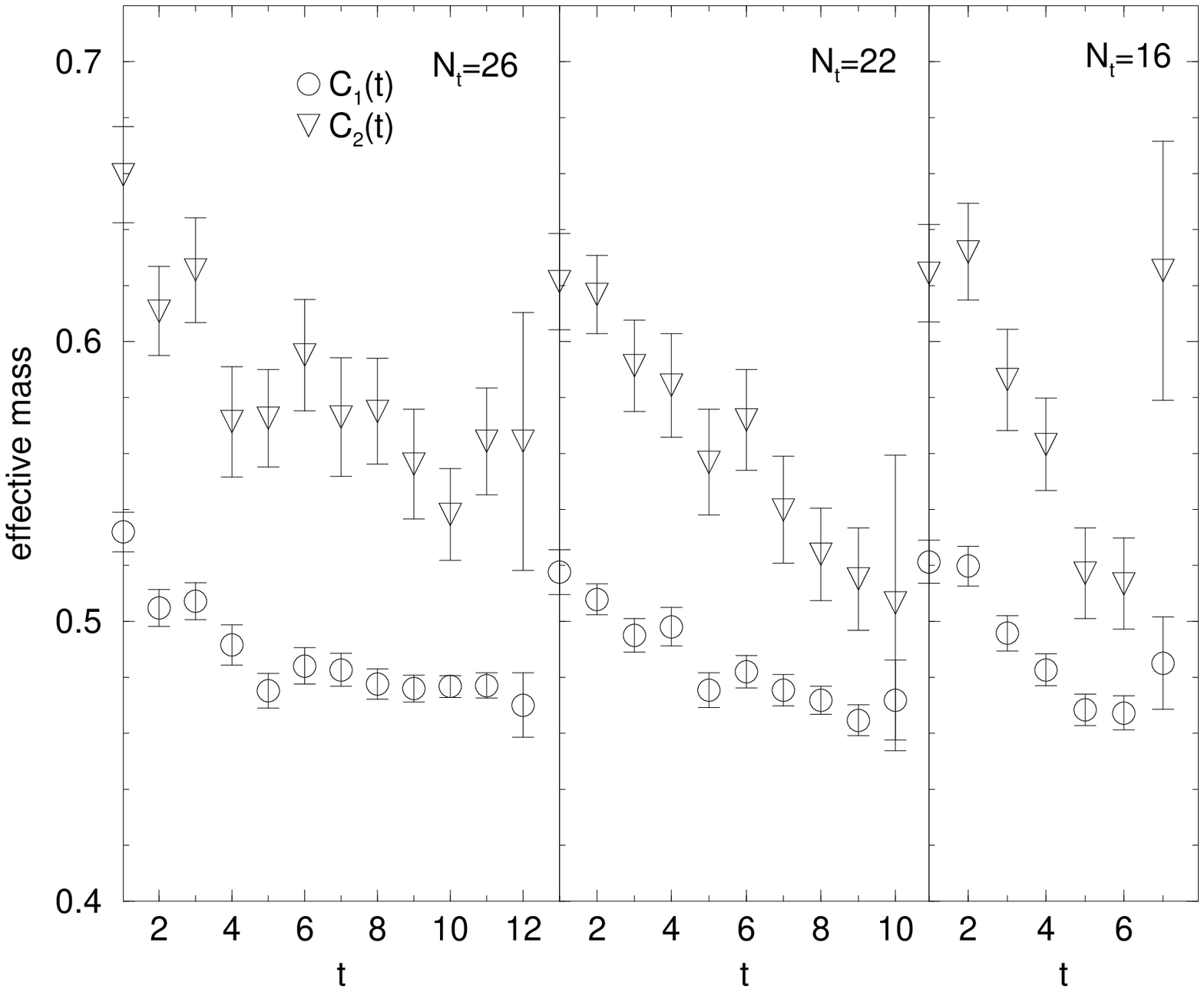,height=2.3in}
\leavevmode\psfig{figure=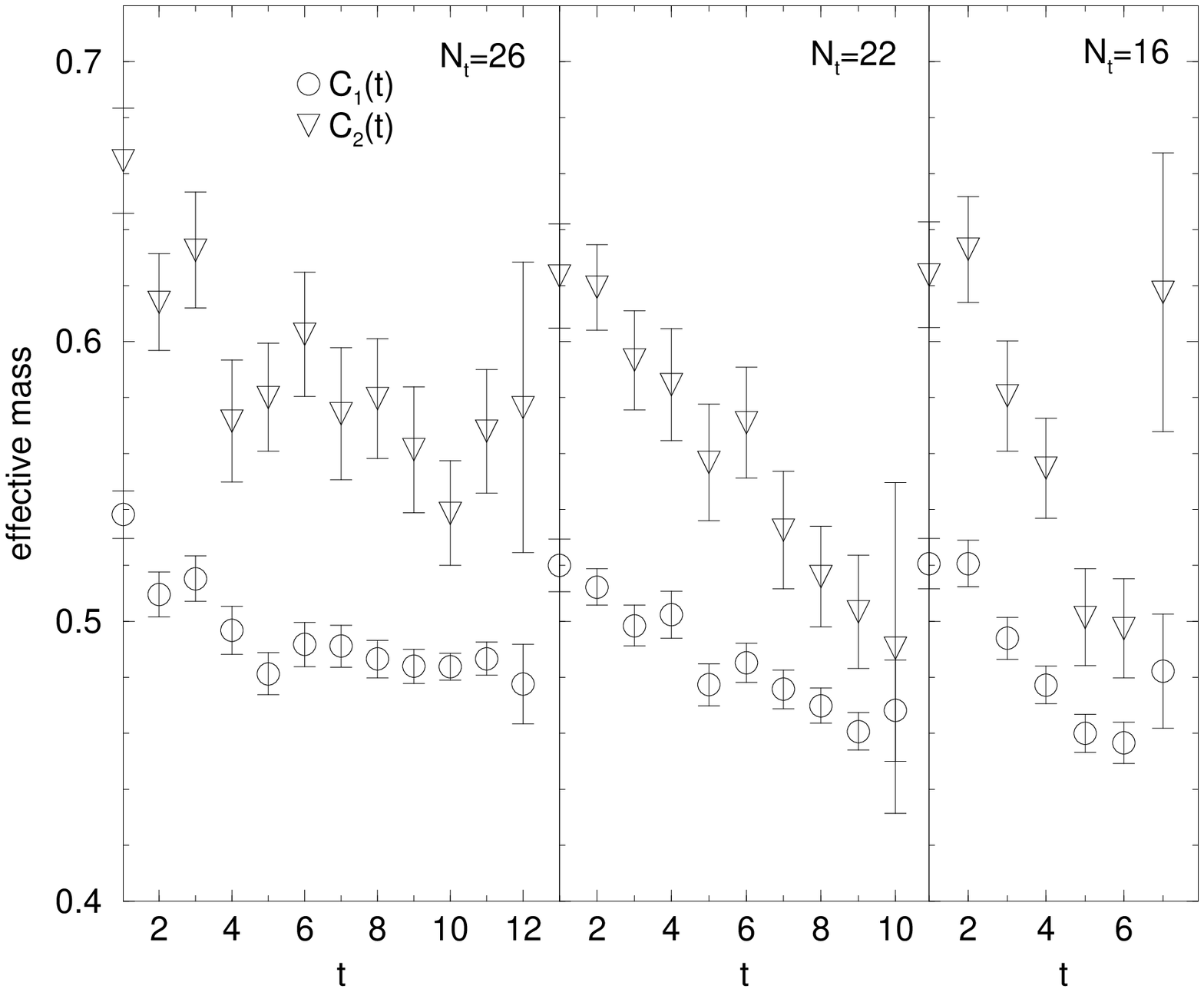,height=2.3in}}
\caption{Effective mass of diagonalized correlator for Set-II at $T>0$.
 Orthogonal matrix at $T=0$ are used for diagonalization.
 The left figure is the result of the pseudoscalar meson and the right one is
 that of the vector meson. Because of the large noises only the lowest two
 effective masses are shown.} 
\label{fig:va_ft}
\end{figure}

In accordance with our strategy the optimized operators which are
constructed at $T=0$ are applied to the case at $T>0$.
The optimized operators are defined from the variational analysis in the 
previous section.
We construct the correlator matrix at $T>0$ with the same basis 
( smearing function ) as $T=0$ ( Fig.~\ref{fig:smear}).
Then the orthogonal matrix defined at $T=0$ are operated to this
correlator matrix, and its diagonal correlators are discussed in this
subsection.
Here we call these correlators as $C_1(t)$, $C_2(t)$ and $C_3(t)$ 
in the order of their effective masses.
Figure~\ref{fig:va_ft} shows the effective masses of $C_1(t)$ and $C_2(t)$
for Set-II, although the effective mass of $C_3(t)$ is too noisy.

Below $T_c$ ( $N_t=26$ ) we can find the
plateau of effective mass. 
The ground and the excited states seem to be observed below $T_c$.
The masses of these states are almost the same as $T=0$ or slightly larger. 
Therefore we conclude that these correlators have little thermal effects
at this temperature, and the spectral structure seems to keep the
form at $T=0$. 
However it is difficult to identify the plateau precisely and determine
the mass quantitatively with the present statistics. More detailed analysis
with higher statistics may open a stage to discuss the potential mass
shift of charmonium near to the $T_c$\cite{c9}.

Above $T_c$ ( $N_t=22$~ and~ $16$ ) 
effective masses have no clear plateau in whole $t$ region.
These behaviors at least signal significant change of correlators when the
system crosses $T_c$.
This behavior appears noticeably in the effective mass of $C_2(t)$. 
In the case of the light quark system investigated in the Ref.~\cite{c8}, 
the effective masses increase as $T$ in the
pseudoscalar and the vector channels.
The observed behavior in present work, however, shows qualitatively
different nature of the correlators.

\begin{figure}[tb]
\center{
\leavevmode\psfig{figure=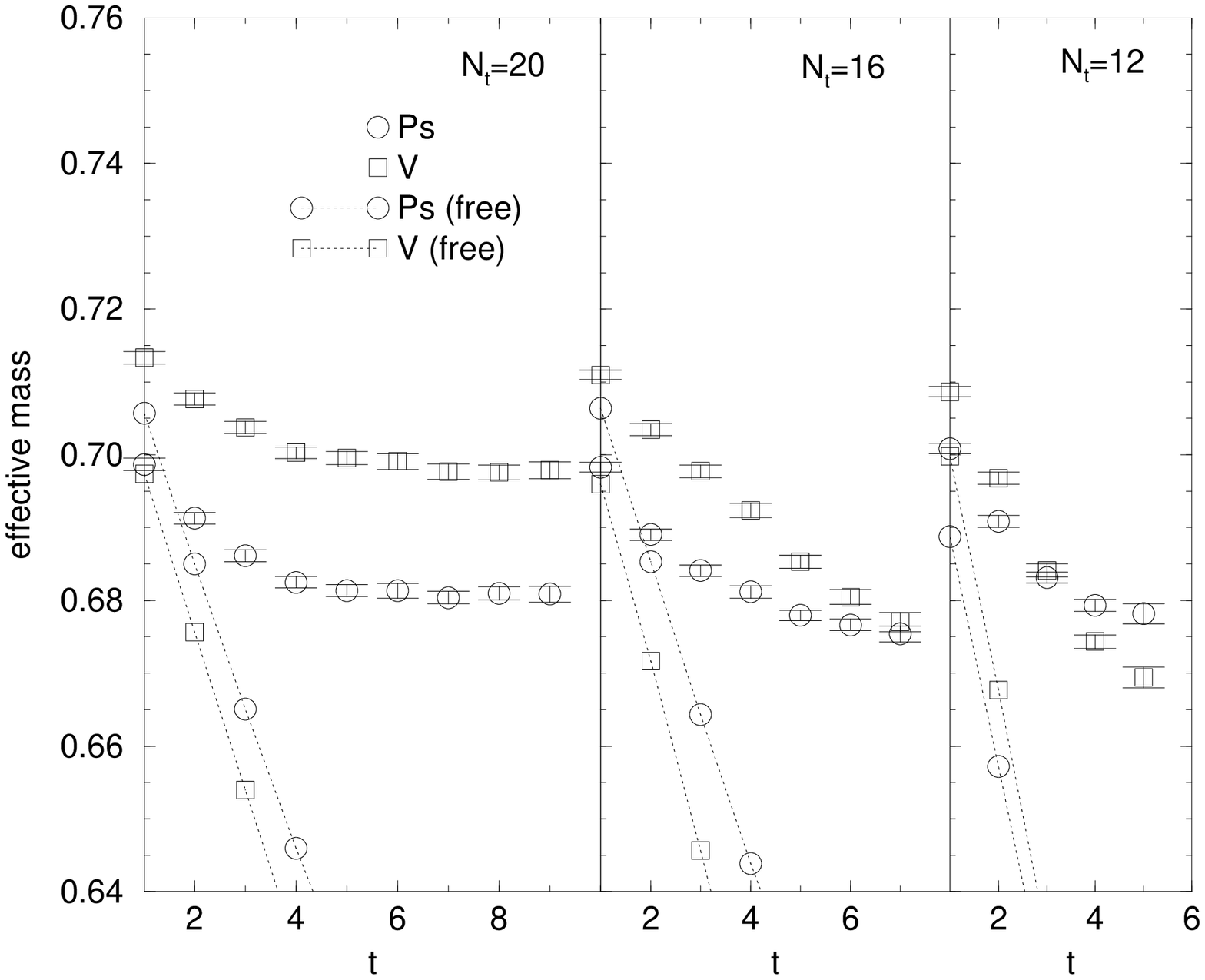,height=2.2in}
\leavevmode\psfig{figure=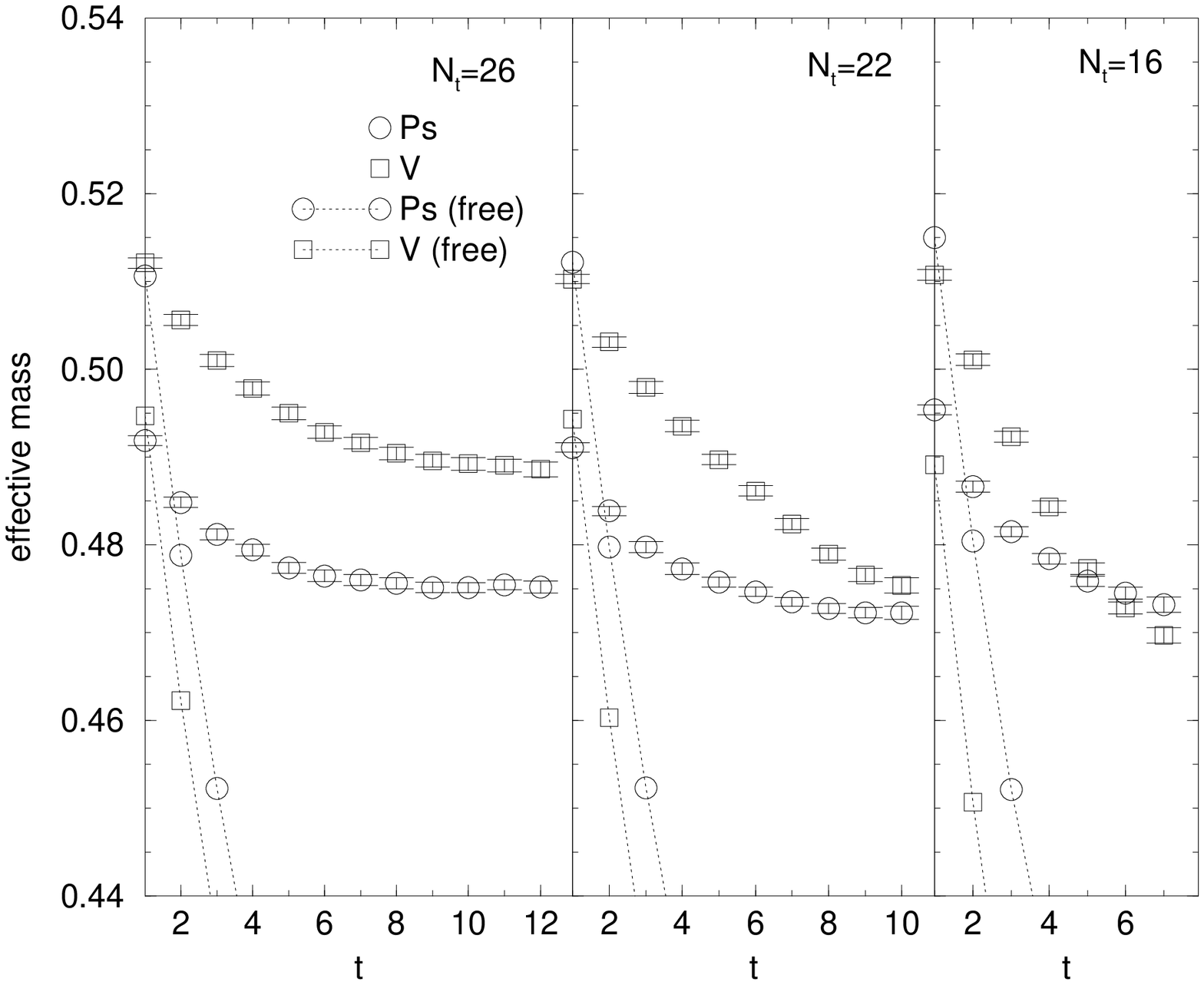,height=2.2in}}
\caption{Effective masses at $T>0$. The left figure is that of Set-I
and the right one is Set-II. In the figures of Set-I, $N_t=20,16$ and 
$12$ correspond to $T/T_c \simeq 0.93, 1.15$ and $1.5$ respectively.
In the figures of Set-II, $N_t=26,22$ and $16$ are $T/T_c=0.93, 1.10$
and $1.53$. The symbols with dotted lines show the case of free quarks.}
\label{fig:ft}
\end{figure}

As the comparison with above results, Fig.~\ref{fig:ft} shows the
effective mass plots at $T>0$ for the 
correlators with smeared source by the wave function and point sink,
described in Sect.~4.1 .

These results are consistent with the variational analysis.
The remarkable change in the vector channel is apparently seen.
Then the order of effective masses for the pseudoscalar and the vector  
mesons is reversed above $T_c$, the same as the case of free quark.
This phenomenon is also brought about in the system of light 
quarks\cite{c8}. 

From the results of this subsection, we observed that the temporal
mesonic correlators change drastically when the system goes through the
deconfining transition.
We can consider the several pictures which can explain (consistent with) the
change for correlators. 
One of these pictures is that the mesonic bound state
disappear in the deconfinement phase.
This is the most interesting case as stated in our motivations.
In order to discuss this picture we investigate the correlation between
$c$ and $\bar{c}$ in the next subsection. 

\subsection{Wave function}
\label{sec:wave}
\vspace*{-0.5pt}
\noindent
In this subsection the $c\bar{c}$ bound state at $T>0$, especially in
the deconfinement phase, is discussed in the light of ``wave function''. 
The definition of the ``wave function'' in the Coulomb gauge is as
follows. 
\begin{equation}
 w_M (\vec{r},t) = \sum_{\vec{x}} \langle
   \bar{q}(\vec{x}+\vec{r},t)\gamma_M
   q(\vec{x},t){\cal O}_M^\dagger(0)\rangle
\end{equation}
Here this definition is the same form as the wave function at $T=0$.  
This wave function shows the spatial correlation between $q$ and
$\bar{q}$, and gives us a hint of the mesonic bound state from its
$t$ dependence. 
In the case of free quarks, $q \bar{q}$ has no bound state, then the
wave function ought to broaden with $t$. 
On the other hand suppose 
quark and anti-quark form a bound state, 
the wave function holds the stable shape with $t$.
We can discuss the existence of such the bound state by observing the
$t$-dependence of the wave function.
For this purpose we compare the correlation at spatial
origin with another spatially separated point at each $t$.
Therefore we define the wave function normalized at the spatial origin,
$\phi_M(\vec{r},t)$, as follows,
\begin{equation}
\phi_M(\vec{r},t) =  
  \frac{w_M(\vec{r},t)}{w_M(\vec{0},t)}.
\end{equation}
From now on the wave function denotes this normalized definition.

Since the question is whether wave function has a stable
shape or not, it is not necessary to use the optimized operator. 
Therefore the smeared source function with exponential form defined as
with Eq.(\ref{eq:exp}),
is used for the analysis of $t$-dependence of the wave function.
\begin{figure}[tb]
 \leavevmode\psfig{figure=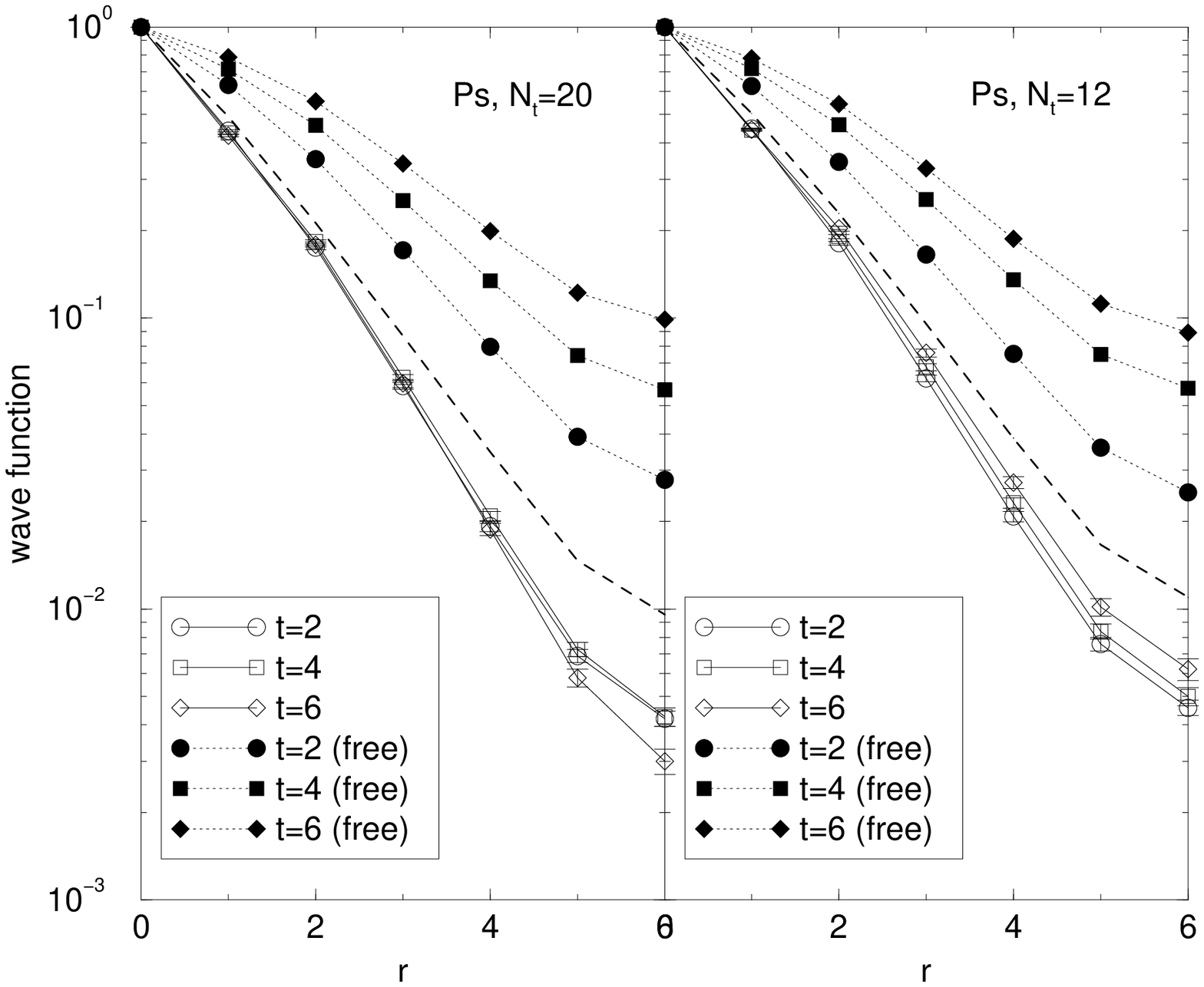,height=2.5in}
 \leavevmode\psfig{figure=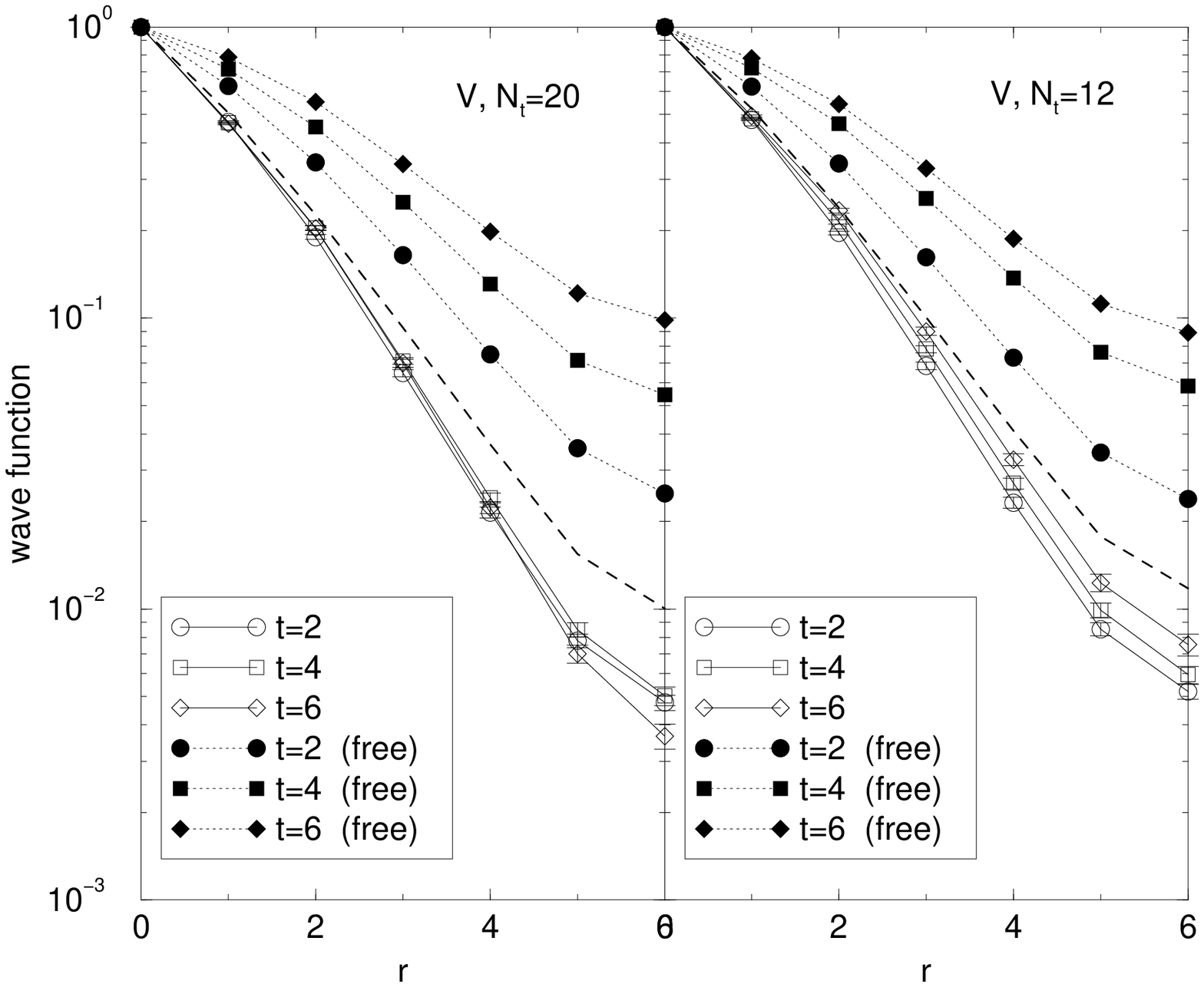,height=2.5in}

 \leavevmode\psfig{figure=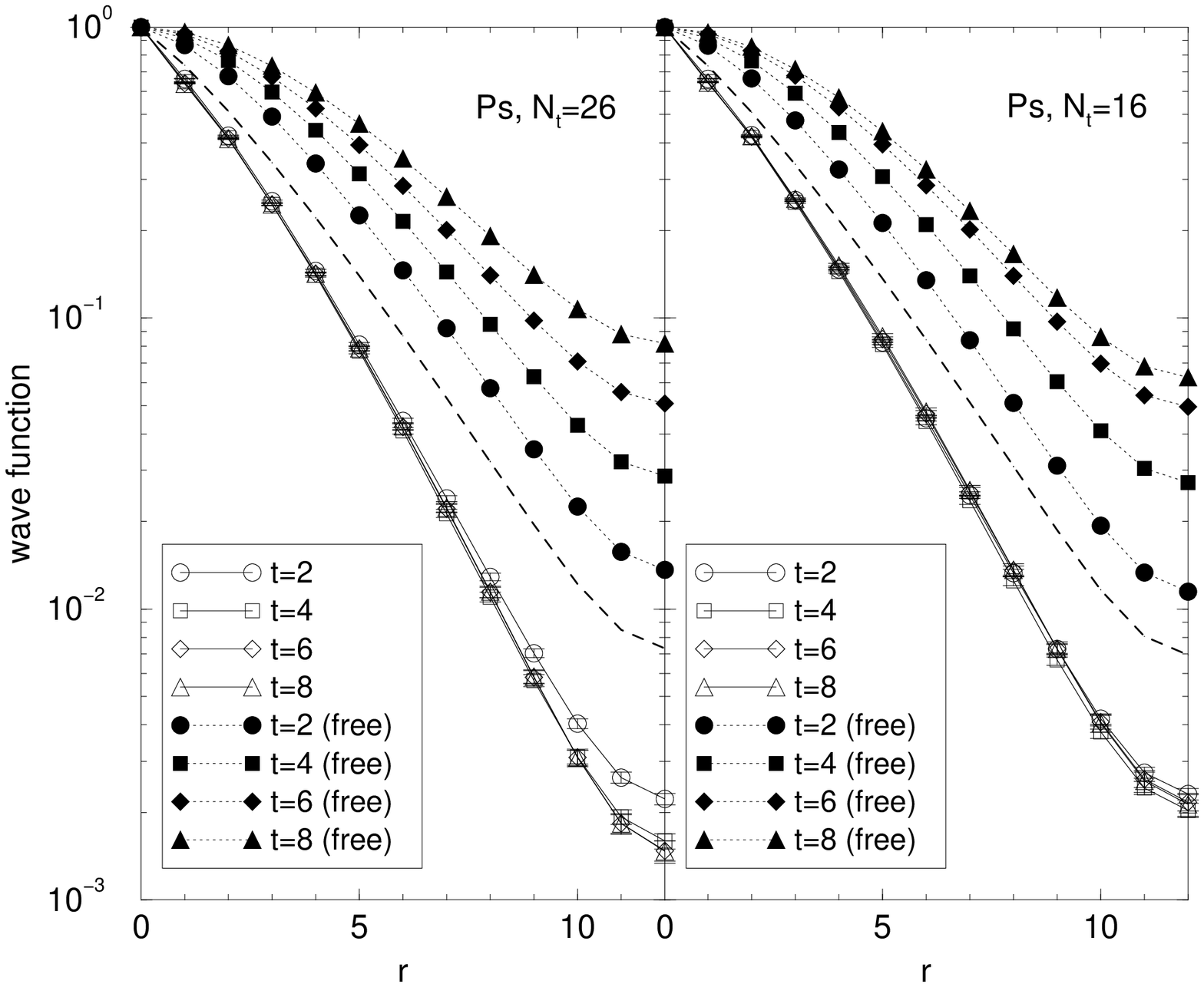,height=2.5in}
 \leavevmode\psfig{figure=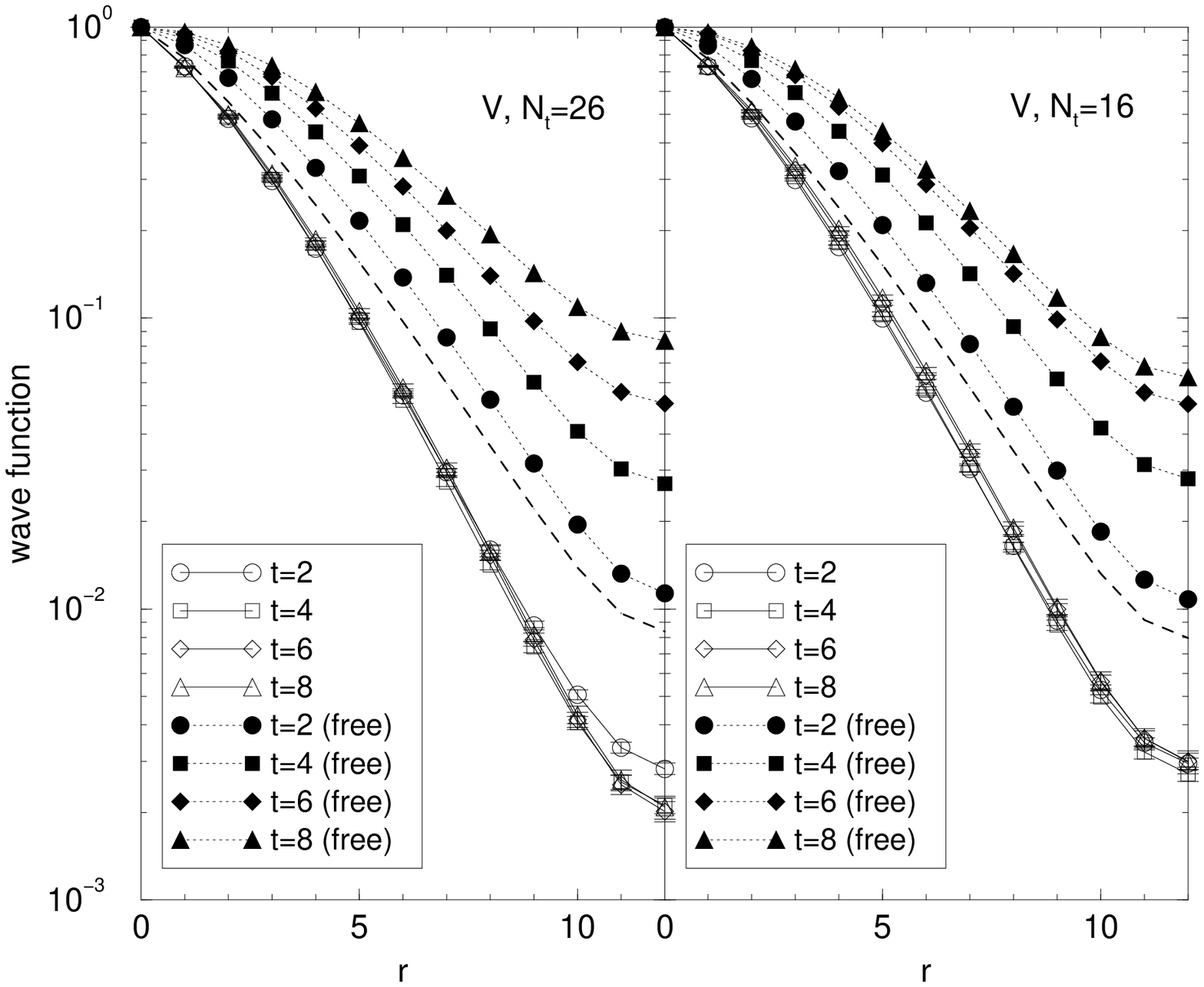,height=2.5in}
\caption{The results of ``wave function'' : dotted lines (full symbols)
 show free quark case, and dashed line (no symbol) show the initial 
shape of ``wave function'' i.e. shape of source function.
 The different symbols correspond to the different $t$ step.
The result of pseudoscalar is left one, and that of vector 
is the other. The top figures are for Set-I and the bottom ones are for
 Set-II. }
\label{fig:wavefunc}
\end{figure}
Fig.~\ref{fig:wavefunc} shows the results at $T>0$
with the smeared source function which is slightly wider than the observed
wave function at $T=0$. 
The wave functions composed of free quark propagators are also shown
together.  

As is shown in the Fig.~\ref{fig:wavefunc}, the behaviors of the
observed wave functions are clearly different 
from that of the free quark case at each temperature and in each mesonic  
channel. In the free quark case the wave functions are broadening as
$t$ as expected. 
On the other hand, the observed wave functions are stable with
the slightly narrower shape than source function.
These behaviors are independent of the source function.

\begin{figure}[tb]
\begin{center}
 \leavevmode\psfig{figure=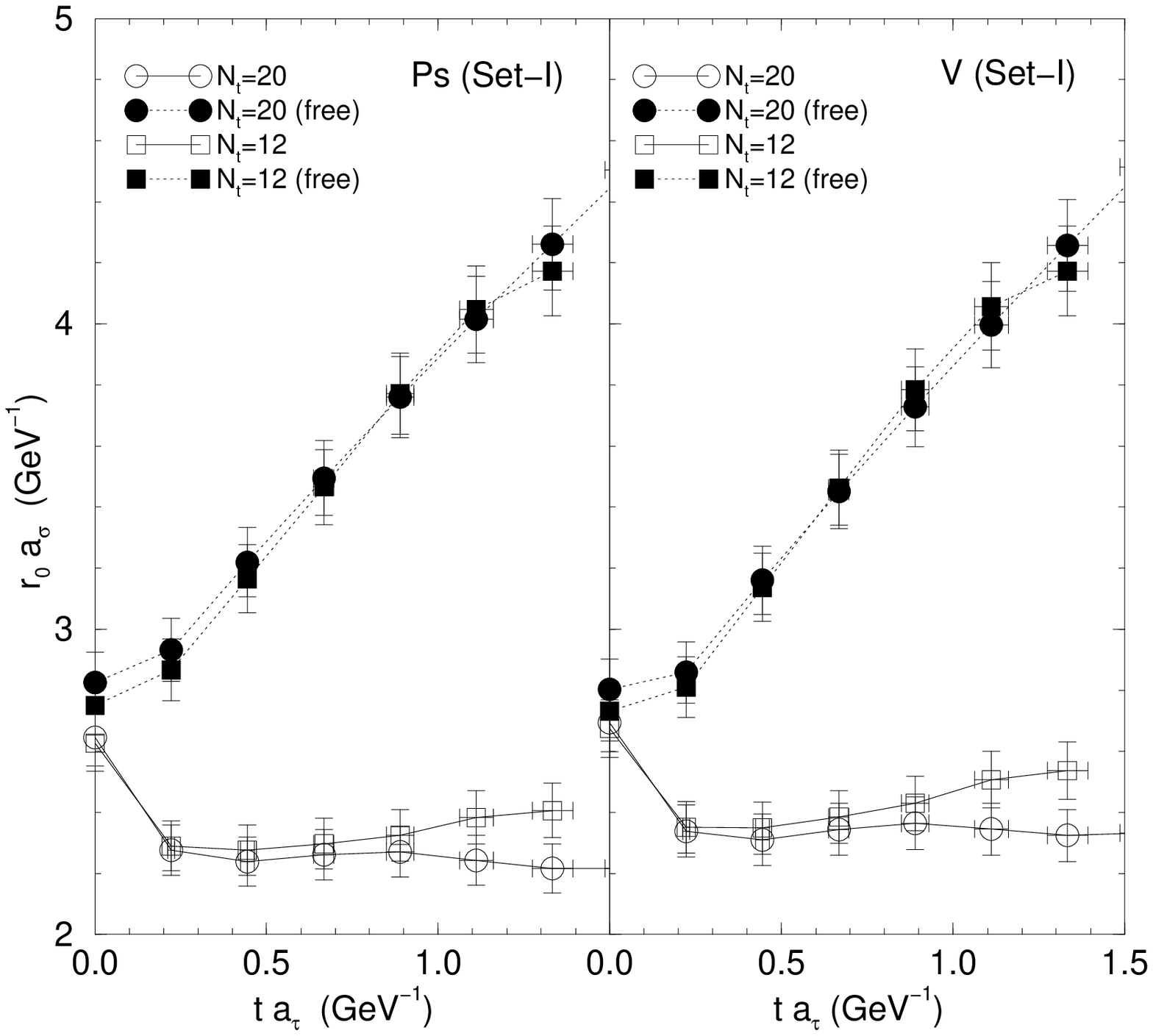,height=2.5in}
 \leavevmode\psfig{figure=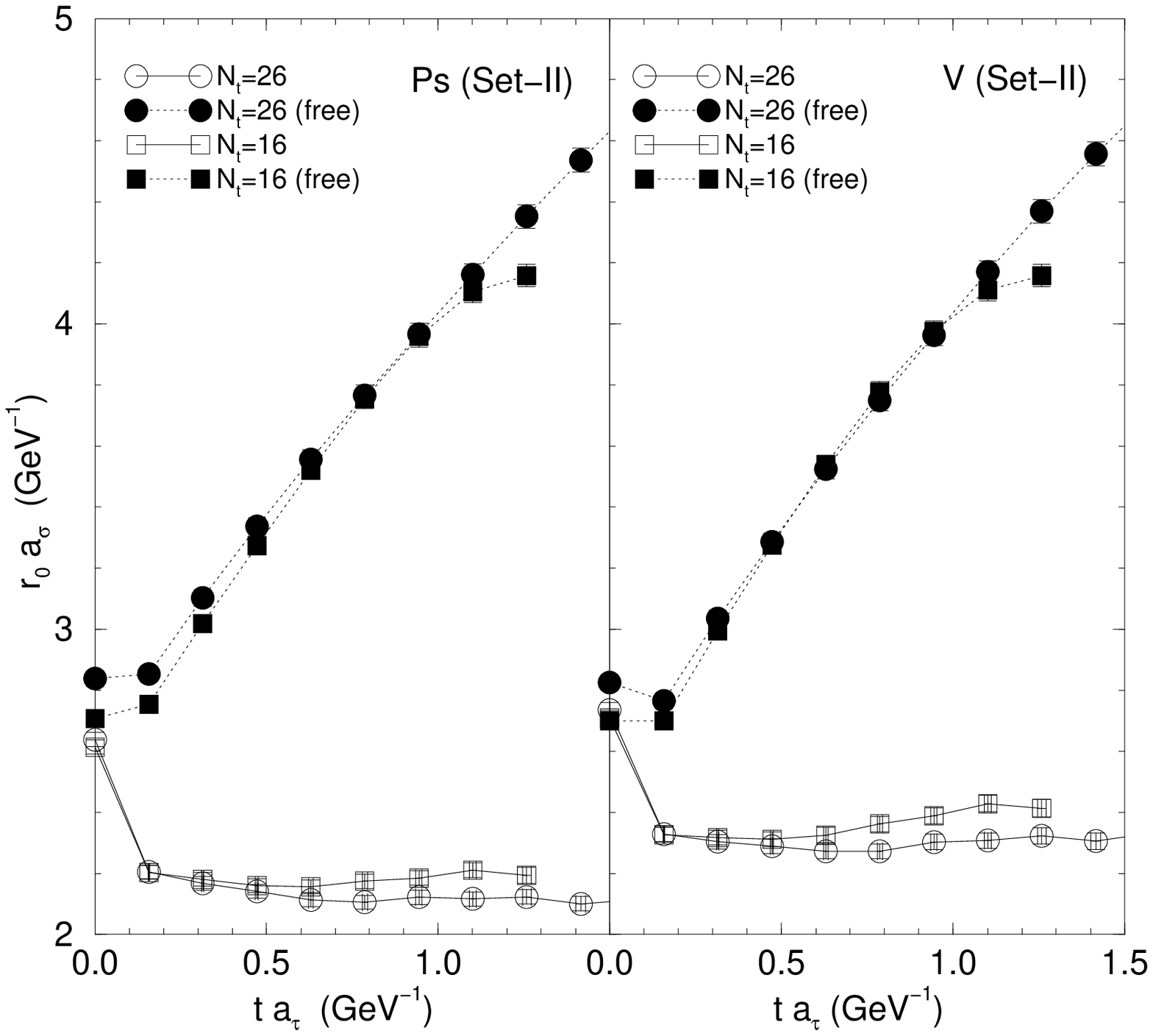,height=2.5in}
\end{center}
\caption{$t$ dependence of $r_0$. Left figure is for Set-I
 and right one is for Set-II.}
\label{fig:r0}
\end{figure}

For a visible expression, 
we define the averaged orbital radius, $r_0$, as 
\begin{equation}
 r_0^3(t)=\sum_z 3 z^2 \phi_M(z,t),
\end{equation}
where we suppose a spherical symmetric wave function and the sum is over
$z$ axis.
These $r_0(t)$ of Set-I and Set-II are shown in Fig.~\ref{fig:r0}.
Figure \ref{fig:r0} shows the $t$ dependence of $r_0$ in the physical
unit, where the error estimation takes into account the error of
$a_{\sigma}$ and $a_{\tau}$.  
These results for Set-I and II are roughly consistent with each other.
The behavior of observed $r_0$ are obviously different from free quark
case, and shows the stable behavior with respect to $t$.
This asymptotic value for the vector channel is larger than that of the
pseudoscalar. 

Below $T_c$, $r_0$ for each channel are almost same as the case of
$T=0$.
At $T\simeq 1.5T_c$ the observed $r_0$ are slightly larger than that of
$T<T_c$.
However the wave function, even in the deconfinement phase, seem to have
the stable $r_0$ independently of the source function.  
Therefore we conclude the same strong spatial correlation as below $T_c$
survives in the deconfinement phase at each mesonic channel. 

The results of this subsection seem to suggest the presence of the hadronic
state in the deconfinement phase, at least up to
$1.5T_c$.
This is the opposite situation to the picture mentioned in the
previous subsection.

\section{Conclusion}
\label{sec:discus}
In this paper we explored the charmonium correlators in the Euclidean
temporal direction at $T>0$ using 
the quenched lattice QCD simulation. 
The high resolution in the temporal direction is achieved by employing the
anisotropic lattice.
We examined the thermal effect on the correlators based on the following
two quantities: the correlators between the optimized operators tuned at
$T=0$, and the $t$-dependence of $q \bar{q}$ wave function.

In the calculation of the wave function in the Coulomb gauge, we found
the 
strong spatial correlation even above $T_c$, up to at least $1.5T_c$.   
These results indicate the quark and the anti-quark tend to be close
each other even in the deconfinement phase.
The similar result is reported in the light quark mass region in
the Ref.~\cite{c8}. 
In the case of charm quark mass these results suggest that the $J/\psi$
may not easily be resolved until $\sim 1.5 T_c$. 
On the other hand, we also observed significant change of the nature of 
the correlators between the operators optimized at $T=0$.
This was signaled by the drastic change of the behavior of effective
mass.
This situation is interesting, and at the same time puzzling.
It is possible to consider several pictures which are able to 
explain our results. 
For example, the mesonic spectral function still have some
peaks above $T_c$, and its width are broadened with thermal effects.
Such a situation naturally explains the observed results in this work.
The existence of hadronic modes just above $T_c$ were also suggested 
by the previous works\cite{c5}.
From the analysis of temporal meson correlators at $T>0$,
the QCD vacuum still has non-perturbative nature above the phase
transition, and far from the perturbative plasma state of quark and gluon. 
In spite of several systematic uncertainties our results in the present
work have important and interesting implications on the fate of hadronic
states above the critical temperature.

Our approach is directly applicable to the dynamical configuration. If
the full QCD simulation on the anisotropic lattice is appropriately
implemented.
Then it is interesting to investigate what effect our results receive from
the dynamical quarks.
To achieve the high resolution in the temporal direction, we adopted the
anisotropic lattice and employed the O(a) improved Wilson quark action 
on it. 
These implementation also useful to study the heavy particle such as the
glueballs\cite{c27} and scalar mesons. 
For the correlators of these states we are inevitably forced to extract 
the signals at the short time separation.
The large temporal lattice cutoff enables us to simulate the heavy particles
on lattices of moderate size with keeping the finite lattice cutoff
effect small. 
The detailed information in the temporal direction is also significantly 
useful for the direct extraction of the spectral function from the lattice
data.
This approach may give us the further information on the spectral structure
of mesons at $T>0$.

\section{Acknowledgments}
We thank the members of QCD-TARO Collaboration for interesting
discussion. 
H.M. thanks T.Onogi, N.Nakajima and J.Harada for useful discussion.
The simulation has been done on 
Intel Paragon XP/S and NEC HSP at the Institute for Nonlinear Science
   and Applied Mathematics, Hiroshima University,
NEC SX-4 at Research Center for Nuclear Physics, Osaka University and
Hitachi SR8000 at KEK.
This work is supported by the Grant-in-Aide for Scientific Research by
Monbusho, Japan ( No. 10640272, No. 11440080 ).
H.M. is supported by the center-of-excellence (COE) program at RCNP,
Osaka University.

\end{document}